\documentclass[a4paper,12pt]{article}

\pdfoutput=1 

\usepackage{jcappub} 
\usepackage[greek,british]{babel}

\usepackage{xcolor}

\usepackage[bottom]{footmisc}

\usepackage{latexsym,array,theorem,mathrsfs,subfigure,bm,float}
\usepackage{amsmath,scalerel}
\usepackage{amsfonts}
\usepackage{amssymb}
\usepackage{xcolor}

\newcommand{\bs}{\boldsymbol}
\newcommand{\bea}{\begin{eqnarray}}
\newcommand{\eea}{\end{eqnarray}}
\newcommand{\be}{\begin{equation}}
\newcommand{\ee}{\end{equation}}
\newcommand{\A}{{\bf a}}
\newcommand{\dA}{\dot{{\bf a}}}

\newcommand{\cc}{{\cal A}}

\newcommand{\xii}{\xi}
\newcommand{\xir}{{\rm r}}
\newcommand{\psii}{\zeta}
\newcommand{\chil}{l}
\newcommand{\yy}{y_0}

\newcommand{\FF}{{$\mbox{{type I FLRW}}$ }}
\newcommand{\FFF}{{$\mbox{type II FLRW}$ }}

\newcommand{\FFd}{{$\mbox{{type I FLRW.}}$ }}
\newcommand{\FFFd}{{$\mbox{type II FLRW.}$ }}

\renewcommand{\theequation}{\arabic{section}.\arabic{equation}}

\begin{document}

\begin{flushright}
{YITP17-14, IPMU17-0033}
\end{flushright}

\title{Anisotropic deformations of spatially open cosmology in massive gravity theory}

\author[a]{Charles Mazuet,}

\emailAdd{\tt charles.mazuet@lmpt.univ-tours.fr}

\affiliation[a]{
Laboratoire de Math\'{e}matiques et Physique Th\'{e}orique CNRS-UMR 7350, \\ 
Universit\'{e} de Tours, Parc de Grandmont, 37200 Tours, France}

\author[b,c]{Shinji Mukohyama,}

\emailAdd{\tt shinji.mukohyama@yukawa.kyoto-u.ac.jp}

\affiliation[b]{Center for Gravitational Physics, Yukawa Institute for Theoretical Physics, 
Kyoto University, 606-8502, Kyoto, Japan}
\affiliation[c]{Kavli Institute for the Physics and Mathematics of the Universe (WPI), 
The University of Tokyo Institutes for Advanced Study, The University of Tokyo, Kashiwa, Chiba 277-8583, Japan}

\author[a,b,d]{Mikhail~S.~Volkov}

\emailAdd{\tt volkov@lmpt.univ-tours.fr}

\affiliation[d]{
Department of General Relativity and Gravitation, Institute of Physics,\\
Kazan Federal University, Kremlevskaya street 18, 420008 Kazan, Russia
}

\abstract{
We study anisotropic deformations of the spatially open 
homogeneous and isotropic cosmology in the ghost free massive gravity theory
with flat reference metric. 
We find that if the initial perturbations are not too strong then 
the physical metric relaxes back to the isotropic de Sitter state. 
However, the dumping of the anisotropies is achieved at the expense 
of exciting  the Stueckelberg fields in such a way that 
the reference metric changes and does not share anymore with the physical metric the same 
rotational and translational symmetries. As a result, the universe evolves 
towards a fixed point which does not coincide with the original solution, 
but for which the physical metric is still de Sitter. 
If the initial perturbation is strong, 
then its evolution generically leads to a singular anisotropic state or, 
for some parameter values, to a decay into flat spacetime. 
We also present an infinite dimensional family of new homogeneous and isotropic cosmologies
in the theory. }

\maketitle

\section{Introduction}
\setcounter{equation}{0}

The main motivation for studying theories with massive gravitons is the fact that they 
offer an explanation for the current 
universe acceleration \cite{1538-3881-116-3-1009,0004-637X-517-2-565}.
Specifically, the ghost-free\footnote{To be precise, the theory is free from the so 
called Boulware-Deser ghost, but it may show other ghosts.} 
massive gravity theory \cite{deRham:2010kj} admits 
self-accelerating cosmological solutions with the Hubble rate 
proportional to the graviton mass.

This theory actually admits infinitely many such vacuum solutions. For all of them the physical metric is de Sitter 
and the reference metric is flat but  the Stueckelberg scalars are different for different solutions. 
There is only one special solution 
for which the physical and reference metrics share the same translational and rotational Killing symmetries
and can be simultaneously diagonalised and put to the standard 
Friedmann-Lema$\hat{\mbox{\i}}$tre-Robertson-Walker 
(FLRW) form \cite{Gumrukcuoglu:2011ew}. 
In what follows we shall call this solution \FFd 
For all other solutions the two metrics share a smaller 
amount of symmetries and cannot be simultaneously brought to the FLRW form
\cite{Chamseddine:2011bu,D'Amico:2011jj,Volkov:2012cf,Volkov:2012zb,Kobayashi:2012fz,Mazuet:2015pea}; 
 we shall call them \FFFd
 
Since both metrics of   \FF solution are simultaneously FLRW, 
the correlation functions of their perturbations are expected to be statistically 
 homogeneous and isotropic.  On the other hand, the correlation functions of perturbations of 
 \FFF solutions are expected to develop statistical inhomogeneity or/and anisotropy, even though  each of the 
two  unperturbed metrics is perfectly FLRW\footnote{For this reason 
 these solutions are sometimes called ``anisotropic FLRW'' or ``inhomogeneous FLRW''}.
 For these reasons  the \FF solution has attracted more attention. 
 
 At the same time, this solution exhibits some peculiar features. First, it is manifestly \FF
 only in the spatially open slicing,  but when its physical metric is expressed in the spatially flat slicing, its 
 reference metric looks inhomogeneous. Secondly, 
 its massive degrees of freedom are not seen within the linear perturbation theory 
 but only at the non-linear level \cite{Gumrukcuoglu:2011zh}. 
 This means that the solution shows  strong coupling, which indicates  that the classical 
 description may break down.   Finally, there are 
 indications that the solution may show  ghost  
 \cite{DeFelice:2012mx,DeFelice:2013awa}. These features, especially the latter one, 
 have been viewed  as obstacles for building realistic cosmology 
 and served a strong motivation for searching for extensions and/or modifications 
 of the original dRGT massive gravity theory. Examples of such modified models 
 that allow for stable self-accelerating de Sitter cosmology include 
 bigravity \cite{Hassan:2011zd,Volkov:2011an,Akrami:2015qga},
extensions
 \cite{DeFelice:2013tsa,Mukohyama:2014rca,DeFelice:2017wel} of the quasidilaton theory
 \cite{DAmico:2012hia,Gabadadze:2014kaa}, generalized massive gravity~\cite{deRham:2014gla}, minimal theory of massive gravity~\cite{DeFelice:2015hla,DeFelice:2015moy,DeFelice:2016ufg}, and so on.

At the same time,  one should emphasise  that Refs.\cite{DeFelice:2012mx,DeFelice:2013awa}
 actually present the stability analysis of a different 
 solution obtained within a different theory and not of the original solution of  Ref.\cite{Gumrukcuoglu:2011ew}. 
 Specifically, Refs.\cite{DeFelice:2012mx,DeFelice:2013awa} consider massive gravity 
 with de Sitter and not flat reference metric, because in such a theory there exists a \FF
 solution with flat spatial sections whose perturbations are relatively easy to study. 
 This solution admits anisotropic generalisations within the Bianchi I class \cite{Gumrukcuoglu:2012aa}\footnote{
 Bianchi I solutions in the theory with anisotropic reference metric have  been studied in \cite{Do:2013tea}.},
 whose analysis has revealed nonlinear ghost instability\footnote{
 At the same type, some of  \FFF solutions turn out to be stable in this case 
 \cite{DeFelice:2013awa}. \label{foot3}}  \cite{DeFelice:2012mx,DeFelice:2013awa}. 
 Now, since this \FF solution of the modified  theory is somewhat similar to the original \FF 
 solution of  Ref.\cite{Gumrukcuoglu:2011ew}, this suggests that the latter may have ghost too. 
However, so far nobody has confirmed or disproved this conjecture  
 by directly studying non-linear deformations of  \FF
 solution of  Ref.\cite{Gumrukcuoglu:2011ew}. 
 Therefore, strictly speaking, the analysis of stability of this solution with a possible detection of ghosts 
 or proving their absence remains an open problem. 
 
 In what follows, as a first step towards our understanding of this problem,
 we shall present our analysis of fully non-linear anisotropic 
 (but homogeneous) deformations of the original \FF  solution  within the Bianchi V class\footnote{
 Anisotropic solutions for all Bianchi types were studied in the bigravity context \cite{Maeda:2013bha}.}.
 In brief, we find that when perturbed, this solution cannot relax back to itself, hence it is unstable. 
 However, if the initial perturbation is not very strong, then the 
 physical de Sitter geometry does relax back to itself and the anisotropies 
 get damped.  During the relaxation the Stueckelberg  fields  change in such a way that the 
reference metric  does not share anymore with the physical metric the same 
rotational and translational symmetries. As a result,  \FF solution evolves 
towards \FFF late time attractor.  
This behavior is similar to what was found in \cite{Gumrukcuoglu:2012aa} in the 
massive gravity with de Sitter reference metric. 
Our analysis does not include perturbations beyond the Bianchi V ansatz and thus the issue of 
ghosts and stability  of \FFF  solutions remain open. (See \cite{DeFelice:2012mx,DeFelice:2013awa}
for the analysis of stability of \FFF solutions in the theory with de Sitter reference metric.)   
We also study  strong initial perturbations and find that their evolution 
generically leads to a singular state where one of the scale factors vanishes. However, 
for some parameter values it may lead to a decay into flat spacetime.

The rest of the text is organised as follows. In the following two Sections we introduce the 
dRGT ghost free massive gravity theory and describe its known homogeneous and isotropic cosmological 
solutions. Section~\ref{IV}  presents the field equations for the anisotropic Bianchi~V metrics.
In Section~\ref{isot} these  equations are analysed for vanishing anisotropies, which yields 
the known type I but  also new 
\FFF solutions. 
In Section~\ref{VI} small anisotropies are studied. Since the first order deviations 
from  \FF solution are trivial  (strong coupling),  we expand up to the second order and find 
that the resulting non-linear equations 
do not admit solutions which tend to zero in the long run. Hence, when perturbed, 
 \FF solution cannot relax to itself. We also analyse linear perturbations around  \FFF solutions 
and find that they all vanish at late times. In Sections~\ref{VII}, \ref{VIII}, and \ref{IX} the anisotropic 
solutions are studied at the fully non-linear level.  Section~\ref{VII} contains the 
analysis of constraints needed to put the equations into the form suitable for numerical integration, 
the equations themselves are displayed in Section~\ref{VIII}, while their numerical solutions 
are described in Section~\ref{IX}. 
A brief summary of our results is given in Section~\ref{X}. 
The special isotropic solutions are considered in Appendix \ref{AppA},
while Appendix \ref{AppB} presents the generalisation of \FFF solutions studied 
in the text to an infinite dimensional family 
of new homogeneous and isotropic dRGT cosmologies. 

We use units in which the length scale is the inverse graviton mass.

\section{The dRGT massive gravity}
\setcounter{equation}{0} 

The theory is  defined on a four-dimensional spacetime manifold endowed with 
two metrics, the physical one $g_{\mu\nu}$ 
and the flat reference metric 
$
f_{\mu\nu}=\eta_{AB}\partial_\mu X^A\partial_\nu X^B
$
with $\eta_{AB}={\rm diag}[-1,1,1,1]$. 
The scalars $X^A(x)$ are sometimes called 
Stueckelberg fields. The theory is defined 
by the action 
\bea                                      \label{1aa}
&&S
=\frac{M_{\rm Pl}^2}{m^2}\int \, \left(
\frac{1}{2}\,R({g})
-{\cal U}\right)\sqrt{-g}\,d^4x
\,,
\eea
where the metrics and all coordinates are 
assumed to be dimensionless, the length scale
being the inverse graviton mass $1/m$.  
 The interaction  between the two metrics 
is determined  by  the tensor 
$
\gamma^\mu_{~\nu}
$
defined by the relation
\be
(\gamma^2)^\mu_{~\nu}\equiv \gamma^\mu_{~\alpha}\gamma^\alpha_{~\nu}={g}^{\mu\alpha}
{f}_{\alpha\nu}. 
\ee
Hence, using the hat to denote matrices, one has
$\hat{\gamma}=\sqrt{\hat{g}^{-1}\hat{f}}$.  
If $\lambda_A$ are eigenvalues of $\hat{\gamma}$
then the interaction potential is
\be
{\cal U}=b_0+\sum_{n=1}^3 b_k\,{\cal U}_k\,,
\ee
where $b_0,b_k$ are parameters and 
${\cal U}_k$ are defined 
by (with 
$[\gamma]\equiv {\rm tr}(\hat{\gamma})$ and 
$[\gamma^k]\equiv {\rm tr}(\hat{\gamma}^k)$) 
\bea                        \label{4}
{\cal U}_1&=&
\sum_{A}\lambda_A=[\gamma],~~~~~
{\cal U}_2=
\sum_{A<B}\lambda_A\lambda_B 
=\frac{1}{2!}([\gamma]^2-[\gamma^2]),\nonumber \\
{\cal U}_3&=&
\sum_{A<B<C}\lambda_A\lambda_B\lambda_C
=
\frac{1}{3!}([\gamma]^3-3[\gamma][\gamma^2]+2[\gamma^3]).
\eea
The metric $g_{\mu\nu}$ and the scalars $X^A$ are the variables of the theory.
Varying the action with respect to $g_{\mu\nu}$
gives the Einstein equations 
\be                \label{einst}
G_{\mu\nu}=T_{\mu\nu}
\ee
with the  energy-momentum tensor 
\bea                                \label{T0}
T^\mu_{~\nu}&=&
\{b_1+b_2\,{\cal U}_1+b_3\,{\cal U}_2
\}\gamma^\mu_{~\nu} 
-\{b_2+b_3\,{\cal U}_1\}(\gamma^2)^\mu_{~\nu} 
+b_3(\gamma^3)^\mu_{~\nu} 
-{\cal U}\,\delta^\mu_\nu \,.~~~~
\eea
Varying with respect to the Stueckelberg fields $X^A$ gives 
the conservation conditions  
\be                              \label{cons}
{\nabla}_\mu T^\mu_{~\nu}=0.
\ee
These equations  also 
follow from the Bianchi identities for the Einstein equations. 

\section{Homogeneous and isotropic cosmologies: a review  \label{II}}
\setcounter{equation}{0}

Equations \eqref{einst} admit a  cosmological solution 
whose physical and reference metrics
are simultaneously homogeneous and isotropic \cite{Gumrukcuoglu:2011ew},
\bea                 \label{Muk}
ds_g^2=-dt^2+\A^2\left[d\rho^2+\sinh^2(\rho) d\Omega^2\right], \nonumber \\
ds_f^2=u_\ast^2\left\{-(d{\A})^2 +\A^2\left[d\rho^2+\sinh^2(\rho) d\Omega^2\right]\right\},
\eea
with $d\Omega^2=d\vartheta^2+\sin^2\vartheta d\varphi^2$ and 
\be
\A=\frac{1}{H}\sinh(Ht).
\ee
Here the Hubble parameter is defined by 
\be
H^2=\frac13(b_0+2b_1 u_\ast+b_2 u_\ast^2),
\ee
where  $u_\ast$  is a root of the algebraic equation 
\be
b_1+2b_2 u_\ast+b_3 u_\ast^2=0. 
\ee
The g-metric is de Sitter  expressed in the open slicing, 
while the f-metric is flat expressed in  Milne coordinates. 
Since both metrics are simultaneously  homogeneous and isotropic, 
we shall call this solution \FFd
The \FF  property is very special and is manifest only in the open slicing, the two metrics 
sharing  the six translational and rotational Killing symmetries associated to this slicing. 
When expressed in spatially flat or closed  slicing, the de Sitter g-metric is still  manifestly 
FLRW but the f-metric looks inhomogeneous because  it does not share the 
corresponding translational symmetries. We shall see this in a moment. 

The theory also admits infinitely many other solutions for which the g-metric 
is de Sitter, but the f-metric cannot be put  to the FLRW form 
simultaneously with the g-metric because the number
of their common symmetries is less than six. 
We shall call such solutions \FFFd
Both type I and \FFF solutions can be described as follows. 
Passing to the  coordinates 
\bea                \label{em1}
x^0=\A\,\cosh(\rho),~~~~
x^1=R\sin\vartheta\cos\varphi,~~~~
x^2=R\sin\vartheta\sin\varphi,~~~~
x^3=R\cos\vartheta,
\eea
with $R=\A\,\sinh(\rho)$,
the f-metric becomes manifestly Minkowski, 
\be                \label{ff}
ds_f^2=u_\ast^2\left\{
-(dx^0)^2+(dx^1)^2+(dx^2)^2+(dx^3)^2
\right\}. 
\ee
Introducing also 
\be
x^4=\frac{1}{H}\cosh(Ht)~~~~\Rightarrow~~~~-(d\A)^2+(dx^4)^2=-dt^2\,,
\ee
the physical metric is 
\be             \label{g2}
ds_g^2=
-(dx^0)^2+(dx^1)^2+(dx^2)^2+(dx^3)^2+(dx^4)^2\,,
\ee
where the coordinates fulfill the relation 
\be               \label{g1}
-(x^0)^2+(x^1)^2+(x^2)^2+(x^3)^2+(x^4)^2=\frac{1}{H^2}.
\ee
This provides the well-known interpretation of de Sitter space as 
4D hyperboloid embedded into 5D Minkowski space.
This parametrisation of the solution 
is convenient for describing more general \FFF solutions.  
For these solutions the g-metric is still described by \eqref{g1},\eqref{g2}
while the f-metric is expressed in terms of the Stueckelberg fields,  
\be                \label{f2}
ds_f^2=
-(dX^0)^2+(dX^1)^2+(dX^2)^2+(dX^3)^2\,,
\ee
where $X^A$ should fulfill equations \eqref{cons}. 
It turns out \cite{Mazuet:2015pea}  that choosing 
\be                \label{ff2}
X^0=u_\ast \,T(x^0 ,x^4),~~~~~~X^k=u_\ast\, x^k\,,
\ee
equations \eqref{cons} reduce to  
\be                   \label{TT}
\left(\frac{\partial T}{\partial x^0}\right)^2-\left(\frac{\partial T}{\partial x^4}\right)^2=1.
\ee
One can obviously choose
 $T=x^0 $ which yields 
  \FF solution. 
However, the PDE admits infinitely many other solutions (they can be constructed explicitly 
\cite{Mazuet:2015pea}), hence the theory admits infinitely many 
\FFF    cosmologies. 
For all these solutions 
the number of common isometries of the two metrics is less than six. 
These solutions may have a peculiar global structure since when 
coordinates $x^0,\ldots x^4$ span the whole of the de Sitter hyperboloid, the 
Stueckelberg fields $X^A$ do not necessarily cover the whole of Minkowski space 
\cite{Motloch:2015gta}.  Examples of other \FFF solutions which are not described by  \eqref{ff2}, \eqref{TT} 
will be given below.    

Let us return for a moment to \FF solution to see how it looks when expressed 
in flat spatial slicing. The coordinates $x^0,x^4$ and $R=\sqrt{(x^1)^2+(x^2)^2+(x^3)^2}$
are then expressed in terms of $\tau,\xir$ as 
\be
Hx^0=\sinh\tau+\frac12\,\xir^2\, a(\tau),~~~~~~
Hx^4=\cosh\tau-\frac12\,\xir^2\, a(\tau),~~~~~~
HR=\xir\, a(\tau), 
\ee
where  $a(\tau)=e^\tau$.
The metrics \eqref{ff} and  \eqref{g2}  become, with $T(\tau,\xir)=x^0$, 
\bea
H^2 ds_g^2=-d\tau^2+a^2(\tau)\left(d\xir^2+\xir^2 d\Omega^2\right), ~~
\frac{H^2}{u_\ast^2}\,ds_f^2=-(dT(\tau,\xir))^2+dR^2+R^2 d\Omega^2. ~~~~~~
\eea
As one can see, the f-metric looks inhomogeneous -- it is not invariant under translations 
of flat slices. This  ``inhomogeneous"  solution had been discovered in \cite{D'Amico:2011jj}
before the solution   \eqref{Muk} was found, and only later it was realised \cite{Mazuet:2015pea} 
that  both  are different forms of the same solution.

\section{Homogeneous and anisotropic cosmologies \label{IV}}
\setcounter{equation}{0}

In what follows we shall be considering
homogeneous and anisotropic cosmologies of the Bianchi V class, 
\bea                              \label{BNVg}
ds_g^2&=&-dt^2+A^2(t)\,dx^2+e^{2x}\left[B^2(t)\,dy^2+C^2(t)\,dz^2\right].
\eea
As we shall see, such metrics can describe anisotropic deformations of the homogeneous 
and isotropic solutions described in the previous Section. 
As we wish the system to be homogeneous, the spatial coordinates $x,y,z$ should 
separate, hence we choose 
the flat reference metric in the form 
\bea                        \label{BNVf}
ds_f^2&=&-(dF)^2+F^2\left[dX^2+e^{2X}(dy^2+dz^2)\right],
\eea
with the Stuckelberg fields 
\bea
F=F(t),~~~~~X=x+f(t).
\eea
One has (here $\bf{A},\bf{B},\bf{C}$ should not be confused with $A,B,C$)
\be                                  \label{gam}
{g^{\mu\sigma}f_{\sigma\nu}}=\left(
\begin{array}{cccc}
{\bf A} & {\bf C}/\Delta & 0 & 0 \\
-\Delta {\bf C}~~~ & {\bf B} & 0 & 0 \\
0 & 0 & {\rm U}_1 & 0 \\
0 & 0 & 0 & {\rm U}_2
\end{array}
\right)\,,
\ee
where 
\bea
{\bf A}&=&\dot{F}^2-F^2\dot{f}^2,~~~~~
{\bf B}=\frac{F^2}{A^2},~~~~~{\bf C}=-\frac{F^2\dot{f}}{A}, \nonumber \\
\Delta&=&\frac{1}{\bf A},~~~~~{\rm U}_1=\frac{F^2}{B^2}\,e^{2f},~~~~~{\rm U}_2=\frac{F^2}{C^2}\,e^{2f}.
\eea
It follows that 
\be                                  \label{gamma}
\bs{\gamma}^\mu_{~\nu}=\sqrt{g^{\mu\sigma}f_{\sigma\nu}}=\left(
\begin{array}{cccc}
a & {c}/({\Delta}) & 0 & 0 \\
-c\Delta & b & 0 & 0 \\
0 & 0 & u_1 & 0 \\
0 & 0 & 0 &  u_2
\end{array}
\right)\,,
\ee
where 
\bea                 \label{a}
a&=&\frac{1}{Y}(\dot{F}^2-F^2\dot{f}^2+{\rm Q}),~~~~~b=\frac{1}{Y}\left(
\frac{F^2}{A^2}+{\rm Q}\right),~~~~~c=-\frac{F^2\dot{f}}{AY},   \nonumber \\
u_1&=&\frac{F}{B}\,e^f,~~~~~u_2=\frac{F}{C}\,e^f\,,
\eea
with
\bea                 \label{b}
Y=\sqrt{\left(\dot{F}+\frac{F}{A}\right)^2 -F^2 \dot{f}^2}\,, ~~~~~~~
{\rm Q}=\frac{F\dot{F}}{A}\,.
\eea
One has 
\be
a+b={Y},~~~~~~~ab+c^2={\rm Q}. 
\ee
Computing the energy-momentum tensor \eqref{T0} gives the following non-trivial components: 
\bea             \label{TTT}
T^0_0&=&-P_0-b\,P_1\,,  \nonumber  \\
T^x_x&=&-P_0-a\,P_1\,,  \nonumber  \\
T^0_x&=&A\,c\,P_1\,, ~~~~~T^x_0=-c\,P_1/A, \nonumber \\
T^y_y&=&-b_0-b_1\,(Y+u_2)-b_2\,[Yu_2+{\rm Q}]-b_3\,u_2\,{\rm Q},  \nonumber  \\
T^z_z&=&-b_0-b_1\,(Y+u_1)-b_2\,[Yu_1+{\rm Q}]-b_3\,u_1\,{\rm Q},  
\eea
where
\be
P_m\equiv b_m+b_{m+1}\,(u_1+u_2)+b_{m+2}\,u_1u_2\,.
\ee
Notice that $T^\mu_\nu$ depends only on time hence the system is indeed homogeneous. 
As a result, 
the Einstein field equations 
$
G^\mu_\nu=T^\mu_\nu
$ 
reduce to a system of five equations for five amplitudes $A,B,C,F,f$. These are three second order equations
\bea                    \label{2}
-\frac{\ddot{C}}{C}-\frac{\dot{C}\dot{B}}{CB}-\frac{\dot{C}\dot{A}}{CA}+\frac{2}{A^2}&=&
\frac12\left(
T^x_x+T^y_y-T^z_z+T^0_0
\right)\,,  \nonumber \\
-\frac{\ddot{B}}{B}-\frac{\dot{C}\dot{B}}{CB}-\frac{\dot{B}\dot{A}}{BA}+\frac{2}{A^2}&=&
\frac12\left(
T^x_x-T^y_y+T^z_z+T^0_0
\right)\,,  \nonumber \\
-\frac{\ddot{A}}{A}-\frac{\dot{C}\dot{A}}{CA}-\frac{\dot{B}\dot{A}}{BA}+\frac{2}{A^2}&=&
\frac12\left(
-T^x_x+T^y_y+T^z_z+T^0_0
\right)\,,  
\eea
and two first order equations 
\bea                 \label{1}
\frac{3}{A^2}
-\frac{ \dot{A}\dot{B} }{AB}
-\frac{\dot{A}\dot{C}}{AC}
-\frac{\dot{B}\dot{C}}{BC}
&=&T^0_0\,,  \nonumber \\
\frac{\dot{C}}{C}+\frac{\dot{B}}{B}-2\,\frac{\dot{A}}{A}&=&T^0_x\,.
\eea
The conservation conditions 
$
\nabla_{\mu}T^\mu_{~\nu}=0
$
reduce to 
\bea              \label{B}
\frac{1}{ABC}\,\frac{d}{dt}\left(ABC\, T^0_0\right)&=&\frac{\dot{A}}{A}\,T^x_x+\frac{\dot{B}}{B}\,T^y_y
+\frac{\dot{C}}{C}\,T^z_z+\frac{2}{A^2}\,T^0_x\,, \nonumber \\
\frac{1}{ABC}\,\frac{d}{dt}\left(ABC\, T^0_x\right)&=&-2\,T^x_x+T^y_y
+T^z_z.
\eea
These can be viewed as equations for the Stuckelberg scalars, because 
they contain the second derivatives $\ddot{F}$ and $\ddot{f}$. 

\subsection{Further reduction}

To simplify the analysis we assume the axial symmetry, 
\be
B=C,
\ee
hence 
\be                  \label{uu}
u_1=u_2=\frac{F}{B}\,e^f\equiv u\,,
\ee
which implies that $T^y_y=T^z_z$. 
The second order Einstein equations \eqref{2} then reduce to 
\bea                  \label{22}
-\frac{\ddot{B}}{B}-\frac{\dot{B}^2}{B^2}-\frac{\dot{A}\dot{B}}{AB}+\frac{2}{A^2}
=&-&P_0-\frac12\,Y P_1\,,     \\
-\frac{\ddot{A}}{A}-\frac{2\dot{A}\dot{B}}{AB}+\frac{2}{A^2}
=&-&P_0 +\left[
u-\frac12\,Y-\frac{F}{AY}\left(\dot{F}+\frac{F}{A}\right) \right]P_1  \nonumber \\
&+&\frac12\left(
Yu-u^2-\frac{F\dot{F}}{A}
\right)dP_1  \,, \nonumber 
\eea
where 
\be
P_m\equiv P_m(u)=b_m+2b_{m+1}\,u+b_{m+2}\,u^2,~~~~~~
dP_m\equiv (P_m(u))^\prime=2(b_{m+1}+b_{m+2}\,u),
\ee
and where we used the fact that 
\be                    \label{iden}
dP_0+u\, dP_1=2P_1.
\ee 
The first order equations \eqref{1} 
reduce to 
\bea             \label{11}
\frac{3}{A^2}-\frac{2\dot{A}\dot{B}}{AB}-\frac{\dot{B}^2}{B^2}&=&-P_0
-\frac{F}{AY}\left(
\dot{F}+\frac{F}{A}
\right)P_1,  \nonumber \\ 
2\,\frac{\dot{B}}{B}-2\,\frac{\dot{A}}{A}&=&-\frac{F^2\dot{f}}{Y}\,P_1\,.
\eea

\section{Isotropic limit    \label{isot}}
\setcounter{equation}{0}
The simplest solutions of the above equations are obtained by setting 
\be
\frac{\dot{A}}{A}=\frac{\dot{B}}{B}. 
\ee
This implies that $A$ and $B$ are proportional to each other, i.e. 
\be                \label{prop}
A=\A,~~~~~~B=e^\chi\, \A,
\ee
with a constant $\chi$. 
Equations \eqref{22},\eqref{11} then reduce to 
\bea                  \label{222}
-\frac{\ddot{\A}}{\A}-\frac{2\dot{\A}^2}{\A^2}+\frac{2}{\A^2}
&=&-P_0-\frac12\,Y P_1\,,   \nonumber   \\
\frac{3}{\A^2}-\frac{3\dot{\A}^2}{\A^2}&=&-P_0
-\frac{F}{\A Y}\left(
\dot{F}+\frac{F}{\A}
\right)P_1,
\eea
and to 
\bea             \label{111}
0
&=&\left[
u-\frac{F}{\A Y}\left(\dot{F}+\frac{F}{\A}\right) \right]P_1
+\frac12\left(
Yu-u^2-\frac{F\dot{F}}{\A}
\right)dP_1  \,,   \nonumber \\ 
0&=&-\frac{F^2\dot{f}}{Y}\,P_1\,.
\eea 
The coefficient $\chi$ in \eqref{prop} 
does not enter these equations, while inserting \eqref{prop} to the line element \eqref{BNVf},
the value of $\chi$ can be changed by a shift $x\to x+ x_0$. Therefore, configurations 
with $\chi\neq 0$ are equivalent to the one  with $\chi=0$. 
It follows that equations \eqref{222} and \eqref{111} describe the isotropic limit.

The second equation in \eqref{111} can be fulfilled by setting either $P_1=0$ or $\dot{f}=0$ or $F=0$. 
In the two latter cases, 
as shown in Appendix \ref{AppA}, solutions of \eqref{222},\eqref{111} describe either flat spacetime 
or configurations with degenerate reference metric. 
Therefore, we choose the $P_1=0$ option  by setting 
\be
u=u_\ast
\ee
where $u_\ast$ is a root of 
\be               \label{PPP}
P_1(u_\ast)=b_1+2 b_2 u_\ast+b_3 u_\ast^2=0. 
\ee
Eqs.\eqref{222} then reduce to 
\bea           \label{eqs}
-\frac{\ddot{\A}}{\A}-\frac{3\dot{\A}^2}{\A^2}+\frac{2}{\A^2}
&=&-P_0(u_\ast)\,,  \nonumber \\
\frac{3}{\A^2}-\frac{3\dot{\A}^2}{\A^2}&=&-P_0(u_\ast), 
\eea
while Eq.\eqref{111} become 
\be               \label{SSS}
Yu_\ast-u_\ast^2-\frac{F\dot{F}}{\A}=0. 
\ee
The first equation in \eqref{eqs} follows from the second one, while the latter can be rewritten as 
\be                      \label{eqA}
\dot{\A}^2-H^2 \A^2=1
\ee
with 
\be                \label{H}
H^2=\frac{P_0(u_\ast)}{3},
\ee 
hence 
\be                  \label{A}
\A=\frac{1}{H}\,\sinh[H (t-t_0)]. 
 \ee
The remaining Eq.\eqref{SSS}  yields 
\be                  \label{FF}
\left(\dot{F}+\frac{F}{\A}\right)^2-F^2\dot{f}^2=\left(
u_\ast+\frac{F\dot{F}}{u_\ast \A}
\right)^2\,,
\ee
whereas  Eq.\eqref{uu} implies that
\be               \label{uast}
u_\ast=\frac{F}{B}\,e^f=\frac{F}{\A}\,e^{f-\chi},
\ee
from which it follows that 
\be                         \label{fff}
\dot{f}=
\frac{\dot{\A}}{\A}-\frac{\dot{F}}{F}. 
\ee
Injecting this to \eqref{FF} 
and setting 
\be
\frac{F}{u_\ast \A}=\sqrt{w}\,, \label{def-w}
\ee
Eq.\eqref{FF} reduces to  
\be                    \label{zzz}
\frac14\,\left(\frac{dw}{d\nu}\right)^2+(w-1)\left(
\frac{dw}{d\nu}+w-\frac{1}{\dot{\A}^2}
\right)=0,
\ee
where $\nu=\ln\A$ and 
 $\dot{\A}^2=1+H^2 e^{2\zeta}$.
Solutions of this equation are 
\be                     \label{w1}
w=1, 
\ee
and also 
\be                  \label{w2}
w=\frac{2 q^2\,\dot{\A}-1-q^4}{H^2\A^2}, 
\ee
where $q$ is an integration constant (notice that $w$ should be positive). 

\subsection{ Type I FLRW solution  \label{sec1} }

Let us first consider the  solution \eqref{w1},  
\be
w=1~~~~\Rightarrow~~~~F=u_\ast\A.
\ee
Eq.\eqref{fff} then implies that $f$ should be a constant
while \eqref{uast} fixes its value, 
\be
f=\chi.
\ee 
Inserting this to \eqref{BNVg},\eqref{BNVf} with $B=C=e^\chi \A$
and performing a shift $x\to x-\chi$ yields 
\bea                              \label{BNV}
ds_g^2&=&-dt^2+\A^2\left(dx^2+e^{2x }\left[dy^2+dz^2\right]\right), \nonumber \\
ds_f^2&=&u_\ast^2\left\{-(d\A)^2+\A^2\left(dx^2+e^{2 x}[dy^2+dz^2]\right)\right\}. 
\eea
This is precisely  the solution \eqref{Muk} because
the spatial parts of the two metrics are both proportional to 
\bea                        
dl^2&=&dx^2+e^{2x}(dy^2+dz^2) \nonumber \\
&=&\frac{1}{\chil^2}\left(
d\chil^2+dr^2+r^2 d\varphi^2
\right)\nonumber \\
&=&d\rho^2+\sinh^2(\rho)[d\vartheta^2+\sin^2\vartheta d\varphi^2], 
\eea
where the coordinates $(x,y,z)$ are related to 
$(\chil,r,\varphi)$ and  next to 
$(\rho,\vartheta,\varphi)$ via 
\be                        \label{cord1}
\chil=e^{-x},~~~~ r e^{i\varphi}=y+iz\,,
\ee
and next 
\be                       \label{cord2}
\cosh(\rho)=\frac{\chil^2+r^2+1}{2\chil},~~~~~\sinh(\rho)e^{i\vartheta}=\frac{\chil^2+r^2-1}{2\chil}+i\,\frac{r}{\chil}. 
\ee
The solutions comprise a two-parameter family.
The first parameter, $u_\ast$, is discrete and takes at most two values since it should fulfill 
the algebraic equation \eqref{PPP} with the additional condition $3H^2=P_0(u_\ast)>0$. The second parameter
is $t_0$ in the definition of $\A$ in \eqref{A}.

\subsection{ Type II FLRW solutions \label{sec2} }

Let us now consider solutions \eqref{w2} for which 
\be                  \label{ww2}
F=\frac{u_\ast}{H}\sqrt{2q^2\,\dot{\A}-1-q^4}, ~~~~~{f-\chi}=\ln \frac{u_\ast \A}{F}.
\ee
Inserting this to \eqref{BNVg},\eqref{BNVf} with $B=C=e^\chi \A$
yields 
\bea                              \label{BNVaa}
ds_g^2&=&-dt^2+\A^2 dx^2+e^{2x }\left[\A^2e^{2\chi}\left[dy^2+dz^2\right]\right],
 \nonumber \\
ds_f^2&=&-dF^2+F^2\left(dX^2+e^{2 X}[dy^2+dz^2]\right), ~~~~~~~~X=x+f(t).
\eea
These solutions comprise a family labeled, apart from $u_\ast$, by three continuous parameters 
$q,\chi$ and $t_0$. 
The g-metric is the same as before and can be transformed to the FLRW form \eqref{Muk}
by absorbing the parameter $\chi$ in the $x$-coordinate. 
However, the same transformation does not bring the f-metric to the FLRW form,
hence these solutions are \FFF. 

These solutions are new and do not belong to the class described by Eqs.\eqref{f2}--\eqref{TT}
in Section \ref{II}. This is indicated already by the fact that for solutions described by \eqref{f2}--\eqref{TT} 
the two metrics share the three rotational symmetries, while for solutions \eqref{BNVaa} 
the common symmetries are the isometries of the $x,y$ space.

As shown in Appendix \ref{AppB}, 
 transforming the f-metric in \eqref{BNVaa}  to the form \eqref{f2} 
and expressing the Stueckelberg fields $X^A$ in terms of coordinates of the 5D Minkowski 
space used in \eqref{g2} gives 
\be                                   \label{B5}
X^0=u_\ast\, \left(x^0+\frac12\,D\right),~~~~X^1=u_\ast\,x^1,~~~~X^2=u_\ast\, x^2,~~~~
X^3=u_\ast\, \left(x^3+\frac12\, D\right),
\ee
with 
\be                         \label{B6}
D=\frac{(Hx^4-q^2)^2}{H^2(x^3-x^0)}. 
\ee
It is also shown in Appendix \ref{AppB} that this 
can be promoted to an infinite dimensional family 
of new \FFF solutions via replacing 
$D$ in \eqref{B5}  by a function that fulfills the non-linear PDE \eqref{BB17}.

\section{Small deviations from isotropy  \label{VI}}
\setcounter{equation}{0}

As we have seen, isotropic solutions in the theory can be either type I
or \FFF described in the previous Section. 
Our next goal is to study slightly anisotropic solutions and we shall therefore consider 
small deformations of the isotropic  backgrounds. 
The principal difference between  type I and \FFF solutions is that the former is 
strongly coupled since its massive degrees of freedom appear only in the second order of perturbation theory,
while the latter admit non-trivial perturbation dynamics at the linear level, at least 
within the Bianchi V class\footnote{It is not known at present 
 if \FFF solutions also have strongly coupled degrees of freedom visible only at the non-linear level.}.
Therefore, when perturbing  \FF solution one is bound to expand up to the second order,
while in \FFF case one can consider only the first order terms \subsection{Perturbations around  \FF   }

Let us assume the configuration to be close to \FF solution, 
\bea                   \label{pertI}
A&=&\A\,(1+\alpha), ~~~~
B=\A\,(1+\beta),  \nonumber \\
\frac{F}{A}&=&u_\ast +\phi, ~~~~~~~~
f=\psi, 
\eea
where the perturbations 
$\alpha,\beta,\phi,\psi$ and their derivatives are small. 
This implies that 
\bea
u=u_\ast +\sigma,  
\eea
with 
\be                      \label{sp}
\sigma=u_\ast\,(\alpha-\beta+\psi)+\phi. 
\ee
One has 
\bea
P_0(u)&=&P_0(u_\ast)+dP_0 (u_\ast)\,\sigma + {\cal O}(\sigma^2),  \nonumber \\
P_1(u)&=&dP_1 (u_\ast)\,\sigma + {\cal O}(\sigma^2), 
\eea
with 
\be
dP_0(u_\ast)=2(b_1+b_2 u_\ast), ~~~~
dP_1(u_\ast)=2(b_2+b_3 u_\ast). 
\ee
Inserting  this to the second order equations \eqref{22}, expanding with respect to the perturbations
and keeping only the leading order terms 
gives equations linear in perturbations, 
%
\bea           \label{p2}
\ddot{\beta}+\frac{\dot{\A}}{\A}\left(5\dot{\beta}+\dot{\alpha}\right)
+\frac{4\alpha}{\A}
=\frac{u_\ast}{2} dP_1(u_\ast) (\dot{\A}-1)\,\sigma, \nonumber \\
\ddot{\alpha}+\frac{\dot{\A}}{\A}\left(2\dot{\beta}+4\dot{\alpha}\right)
+\frac{4\alpha}{\A}
=\frac{u_\ast}{2} dP_1(u_\ast) (\dot{\A}-1)\,\phi.
\eea
Expanding similarly the first order equations \eqref{11} gives 
\bea          \label{p10}
2\frac{\dot{\A} }{\A}\left(\dot{\alpha}+2\dot{\beta}\right)+\frac{6\alpha}{\A^2}&=&0,  \nonumber \\
2(\dot{\alpha}-\dot{\beta})&=&0. 
\eea
The second of these equations implies that $\dot{\beta}=\dot{\alpha}$ while the first one reduces then to 
\be                       \label{al0}
\dot{\alpha}=-\frac{\alpha}{\A\dot{\A}}\,.
\ee
As a result, 
the left hand sides of the two equations \eqref{p2} reduce to the same expression, 
\be
\ddot{\alpha}+\frac{6\dot{\A}}{\A}\,\dot{\alpha}+\frac{4}{\A}\,\alpha
=\frac{\A\ddot{\A}-\dot{\A}^2+1}{\A^2\dot{\A}^2}\,\alpha=0,
\ee
where we used the equations for the background $\A$. 
Therefore, the right hand sides of Eqs.\eqref{p2} vanish, hence $\sigma=\phi=0$. 
Eq.\eqref{sp} implies in this case that $\psi=\beta-\alpha$ is a constant whose value can be 
set to zero by redefining the $x$-coordinate. This gives $\alpha=\beta$. 
Eq.\eqref{al0} implies that 
\be
\frac{d\alpha}{d\A}=-\frac{\alpha}{\A\dot{\A}^2}=-\frac{\alpha}{\A\,(1+H^2\A^2)}
~~~~\Rightarrow~~~~\alpha=const.\times \frac{\dot{\A}}{\A}.
\ee
As a result, one has $\delta A=\delta B=const.\times \dot{\A}$ and this
corresponds to the 
change of the background solution induced by shifting the reference time moment $t_0$ in \eqref{A}.

Therefore, the dynamics of linear perturbations  around  \FF  background is trivial. 
In order to obtain 
something non-trivial, one has to expand the right hand sides of Eqs.\eqref{11} up to 
second order terms, which gives 
\bea          \label{p1}
2\frac{\dot{\A} }{\A}\left(\dot{\alpha}+2\dot{\beta}\right)+\frac{6\alpha}{\A^2}&=&dP_1(u_\ast)\,\sigma\,(\phi+\frac12\,\sigma),  \nonumber \\
2(\dot{\alpha}-\dot{\beta})&=&dP_1(u_\ast)\frac{\A^2}{\dot{\A}+1}\,\sigma\,[\dot{\sigma}-\dot{\phi}
+u_\ast(\dot{\beta}-\dot{\alpha})]. 
\eea
On the right one can neglect the cubic and higher order terms since they are 
subdominant as compared  to the quadratic terms. As a result, equations \eqref{p1} contain both on the left 
and on the right only  terms leading in perturbations. The equations 
can be resolved with respect to $\dot{\alpha}$ and $\dot{\beta}$,
\bea                 \label{ab0}
\dot{\alpha}=-\frac{\alpha}{\A\dot{\A}}+{\cal S}_\alpha\,,~~~~~
\dot{\beta}=\dot{\alpha}+{\cal S}_\beta\,,\, 
\eea
with 
\be
{\cal S}_\beta=\frac{dP_1 \A^2}{\cal N}(\dot{\phi}-\dot{\sigma})\sigma,~~~~~
{\cal S}_\alpha=\frac{dP_1 \A}{12\dot{\A} }(\sigma+2\phi)\sigma-\frac{2}{3}S_\beta,
\ee
where
\be
{\cal N}=dP_1 u_\ast \A^2\sigma+2\dot{\A}+2.
\ee
Injecting everything  to Eqs.\eqref{p2} gives a closed system of two equations for $\sigma,\phi$,
\bea
\dot{\cal S}_\alpha+\dot{\cal S}_\beta-\frac{1}{\A\dot{\A}}\,{\cal S}_\alpha+
\frac{\dot{\A}}{\A}\left(6{\cal S}_\alpha+5{\cal S}_\beta\right)&=&
\frac{u_\ast}{2}\,dP_1(u_\ast)(\dot{\A}-1)\sigma\,,\nonumber \\
\dot{\cal S}_\alpha-\frac{1}{\A\dot{\A}}\,{\cal S}_\alpha+
\frac{\dot{\A}}{\A}\left(6{\cal S}_\alpha+2{\cal S}_\beta\right)&=&\frac{u_\ast}{2}\,dP_1(u_\ast)(\dot{\A}-1)\phi\,.  \label{SS}
\eea
These equations simplify for $\A\gg 1$ since one has in this case 
\be
\dot{\A}=\sqrt{1+H^2\A^2}\approx H\A,~~~~~~\dot{\A}\pm 1\approx H\A,
\ee
hence 
\be
{\cal N}=dP_1 u_\ast \A^2\sigma+2\dot{\A}+2\approx (dP_1 u_\ast \A\sigma+2 H)\A\approx 2H\A. 
\ee
Here the second approximation is implied by the first equation in \eqref{p1}, whose 
 left hand side is small and hence the right hand side 
 proportional to $u_\ast dP_1 {\A}\sigma$ should be small too. 
As a result,
\be
{\cal S}_\beta\approx \frac{dP_1(u_\ast) \A}{2H}(\dot{\phi}-\dot{\sigma})\sigma,~~~~~
{\cal S}_\alpha\approx \frac{dP_1(u_\ast) }{12 H}(\sigma+2\phi)\sigma-\frac{2}{3}S_\beta\,.
\ee
Inserting this to \eqref{SS} with the small terms neglected, 
\bea
\dot{\cal S}_\alpha+\dot{\cal S}_\beta+
H\left(6{\cal S}_\alpha+5{\cal S}_\beta\right)&=&\frac{u_\ast}{2}\,dP_1(u_\ast) H\A\sigma\,,\nonumber \\
\dot{\cal S}_\alpha+
H\left(6{\cal S}_\alpha+2{\cal S}_\beta\right)&=&\frac{u_\ast}{2}\,dP_1(u_\ast) H\A\phi\,,  \label{SSa}
\eea
yields 
\bea
\left[\sigma\left(\frac12\sigma+\phi+\A\,(\dot{\phi}-\dot{\sigma})\right)\right]^{\mbox{.}}
+3H\sigma(\sigma+2\phi+\A(\dot{\phi}-\dot{\sigma}))&=&3H^2u_\ast \A\sigma,  \nonumber \\
\left[\sigma\left(\frac12\sigma+\phi-2\A\,(\dot{\phi}-\dot{\sigma})\right)\right]^{\mbox{.}}
+3H\sigma(\sigma+2\phi-2\A(\dot{\phi}-\dot{\sigma}))&=&3H^2u_\ast \A\phi\,.
\eea
Expressing the perturbations as 
\be
\sigma=\frac{W+Z}{3},~~~~~~\phi=\frac{W-2Z}{3},
\ee
these equations reduce to 
\bea                                 \label{pert}
\left((W+Z)\dot{Z}\right)^{\mbox{.}}+4H(W+Z)\dot{Z}+3u_\ast H^2 Z=0, \nonumber \\
{W\dot{W}-Z\dot{Z}+3H(W^2-Z^2)=3u_\ast H^2\,\A\,W.   }
\eea
These equations have been derived assuming the perturbations and their derivatives to be small. 
Therefore, only those solutions make sense for which $W,Z$ and their derivatives are small. 
Let us assume $W,Z,\dot{W},\dot{Z}$ to be small. 
The second equation in \eqref{pert} is  
\be
W(\dot{W}+3HW-3u_\ast H^2\A)=Z\dot{Z}+3HZ^2,
\ee
and since $\dot{W}$ and $HW$ are small, they can be neglected 
as compared to the large term $u_\ast H^2\A$,
hence
\be
W\approx -\frac{Z\dot{Z}+3H^2Z^2}{3u_\ast H^2\A}. 
\ee
Next, one has 
\be
W+Z\approx \left(1-\frac{\dot{Z}+3HZ}{3u_\ast H^2\A}\right)Z\approx Z,
\ee
since $\dot{Z},Z$ are small, therefore the first equation in \eqref{pert} reduces to 
\bea                                 \label{WZ1}
\left(Z\dot{Z}\right)^{\mbox{.}}+4HZ\dot{Z}+3u_\ast H^2 Z=0.
\eea
Setting 
$\dot{Z}=p(Z)$ transforms  this equation to 
\be
Zp\,\frac{dp}{dZ}+p^2+HZ(4p+3u_\ast H)=0,
\ee
and since $p=\dot{Z}$ is small by assumption, one has $4p\ll 3u_\ast H$, hence the equation 
can be replaced by 
\be
Zp\,\frac{dp}{dZ}+p^2+3u_\ast H^2 Z=0. 
\ee
This can be integrated  to give 
\be
p=\dot{Z}=\sqrt{\frac{\tilde{C}}{Z^2}-2u_\ast H^2 Z  }. 
\ee
Now, if the integration constant $\tilde{C}\neq0$, then $\dot{Z}\to\infty$ as $Z\to 0$, which  would 
contradict the assumption of smallness of derivatives. 
Hence one has to set $\tilde{C}=0$, which finally gives the solution, 
\be                            \label{WZ}
Z=-\frac{u_\ast H^2}{2}\, (t-t_\ast)^2,~~~~~~W=\frac{u_\ast H^2}{2\A}\,(t-t_\ast)^3,
\ee
where $t_\ast$ is another integration constant. 
This is the most general solution of Eqs.\eqref{pert} for which $W,Z$ and their first derivatives   
are small. However, they are small only in the vicinity of $t= t_\ast$ and 
diverge for $t\to\infty$, hence they
cannot approach zero  asymptotically. 
Therefore, {when perturbed,  \FF solution cannot relax back to itself 
in the long run}.  It follows that the anisotropic  configuration must either  oscillate around 
the unperturbed \FF background, 
or approach some other background for $t\to\infty$, or  hit a singularity 
at some point. The latter two options are confirmed by the  numerical analysis. 

The existence of the solution \eqref{WZ} actually indicates that the standard formulation of the Cauchy 
problem should be modified when applied to type I background. Indeed, 
the functions $W$ and $Z$ vanish at $t=t_\ast$ together with their first derivatives   
but differ from zero for $t\neq t_\ast$. 
There is also the  solution for which $Z=W=0$ everywhere, in particular at $t=t_\ast$. 
Therefore,
specifying the functions and their first derivatives at $t=t_\ast$ does not specify the solution uniquely. 
From the mathematical viewpoint this simply means that $Z=W=0$ is a singular point of 
differential equations, in which case the solution is not necessarily specified by values of $Z,W$
and their first derivatives, but maybe by their second and higher derivatives. This does not mean 
that the predictability is lost but rather shows that the standard formulation of the Cauchy problem 
should be modified when applied to \FF background 
(see \cite{Motloch:2016msa} for discussion of other  difficulties  of the Cauchy 
analysis in massive gravity).

\subsection{Perturbations of  \FFF   }

Let us now assume the configuration to be close to one of \FFF solutions, 
\bea
A&=&\A\,(1+\alpha),  ~~~~~~
B=\A\,(1+\beta),  \nonumber \\
F&=&u_\ast\A\sqrt{w}\, (1 +\phi),  ~~~~~~
u=u_\ast +\sigma,  
\eea
where 
$\alpha,\beta,\phi,\sigma$ are small. 
One has 
\be
\dot{f}=-\frac{\dot{w}}{2w}+\frac{\dot{\sigma}}{u_\ast}+\dot{\beta}-\dot{\phi}+\ldots 
\ee
where the dots denote terms non-linear in perturbations. 
Expanding equations \eqref{22} and \eqref{11} one finds that both their left-hand and right-hand sides 
contain terms linear in perturbations. 
Let us first notice that those parts of the first equation in \eqref{22} and of the two equations \eqref{11} that are linear in perturbations comprise a closed subsystem of three equations, 
 \bea           \label{pp2}
\ddot{\beta}+\frac{\dot{\A}}{\A}\left(5\dot{\beta}+\dot{\alpha}\right)+\frac{4}{\A^2}\,\alpha&=&\left(
dP_0+\frac12\,Y dP_1
\right)\sigma ,  \nonumber \\
2\,\frac{\dot{\A}}{\A}\left(\dot{\alpha}+2\dot{\beta}\right)+\frac{6}{\A}\,\alpha&=&\left(
dP_0 +\frac{F}{\A Y}\left( \dot{F}+\frac{F}{\A} \right)dP_1
\right)\sigma , \nonumber \\
2(\dot{\alpha}-\dot{\beta})&=&\frac{F^2\dot{f}}{Y}\,dP_1 \,\sigma ,
\eea
where $\A,F,Y,dP_0,dP_1$ correspond to the background solution \eqref{ww2}. 
The last two of these equations can be resolved with respect to $\dot{\alpha}$ and $\dot{\beta}$,
\bea                 \label{ab}
\dot{\alpha}=-\frac{\alpha}{\A\dot{\A}}+S_\alpha\,\sigma,~~~~~
\dot{\beta}=\dot{\alpha}+S_\beta\,\sigma,\, 
\eea
with 
$$
S_\alpha=\frac{dP_1}{6Y\A\dot{\A}}\left((2\dA^2+1)F^2-2\A\dA F\dot{F}-u_\ast Y\A^2
+\A F\dot{F}\right), ~~~
S_\beta=\frac{dP_1 F(\A\dot{F}-F\dA)}{2Y\A}. 
$$
Injecting $\dot{\alpha}$ and $\dot{\beta}$ into the first equation in \eqref{pp2} yields 
a first order equation for $\sigma$,
\be
\dot{\sigma}+\frac{4H^2\A}{\dA-q^2}\,\sigma=0~~~~\Rightarrow~~~~
\sigma=C_\sigma H^4\exp\left(\int_t^\infty  \frac{4H^2\A}{\dA-q^2}\,d t   \right),
\ee
where $C_\sigma$ is an integration constant. 
Injecting this to \eqref{ab} and integrating gives 
\be
\alpha=\frac{\alpha_\infty}{H}\,\frac{\dA}{\A}-\frac{\dA}{\A}\int^\infty_t \frac{\A}{\dA}\,S_\alpha\,\sigma\,dt,
~~~~~~\beta=\beta_\infty-\alpha_\infty+\alpha-\int^\infty_t S_\beta\, \sigma dt,
\ee
where $\alpha_\infty$ and $\beta_\infty$ are integration constants.  
One has at late times for $\A\to\infty$  
\bea
\sigma~~&\to&~~
\frac{C_\sigma}{\A^4}\left(1+{\cal O}\left(\A^{-1}\right)\right), \\
\alpha~~&\to&~~\alpha_\infty\left(1+ \frac{1}{2H^2\A^2}+{\cal O}\left(\A^{-4}\right)\right)
-C_\sigma\left(
\frac{u_\ast q^2 dP_1}{9H(q^2+1)\,\A^3}+{\cal O}\left(\A^{-4}\right)
\right),    \nonumber   \\
\beta~~&\to&~~\beta_\infty+\alpha_\infty\left(\frac{1}{2H^2\A^2}+{\cal O}\left(\A^{-4}\right)\right)
+C_\sigma\left(
\frac{u_\ast q^2 dP_1}{18H(q^2+1)\,\A^3}+{\cal O}\left(\A^{-4}\right)
\right).   \nonumber   
\eea
Let us finally linearise the second equation in \eqref{22}, 
\be
-\ddot{\alpha}-2\,\frac{\dA}{\A}\left(2\dot{\alpha}+\dot{\beta}\right)-\frac{4}{\A^2}\,\alpha=
\left[
u_\ast -\frac{F}{\A Y}\left(\dot{F}+\frac{F}{\A}\right) \right]dP_1\, \sigma
+\frac{dP_1}{2}\,\delta \left(
Yu_\ast -\frac{F\dot{F}}{A}
\right), \nonumber 
\ee
where $\delta$ denotes the linear in perturbations part.  
Using the above equations for $\alpha,\beta,\sigma$, 
this equation reduces to 
\be               \label{phi}
\dot{\phi}-\frac{\dot{\yy}}{\yy}\,\phi=\Sigma_\alpha \alpha+\Sigma_\sigma \sigma.
\ee
Here one has 
 at late times $\Sigma_\sigma={\cal O}(\A)$ and 
$\Sigma_\alpha={\cal O}(\A^{-1})$ while 
$\yy(t)$ is obtained by varying the background amplitude $F$ with respect to the parameter 
 $q$,
\be
\yy(t)=\frac{1}{F}\frac{dF}{dq}. 
\ee
The solution of \eqref{phi} is 
\be
\phi=\phi_\infty \,\yy(t)-y\int_t^\infty \frac{dt}{y}\,(\Sigma_\alpha \alpha+\Sigma_\sigma \sigma),
\ee
where $\phi_\infty$ is yet another integration constant. One has at late times 
\bea
\phi\to&\phi_\infty&\left(
1+\frac{1-q^4}{2Hq^2\,\A}+{\cal O}(\A^{-2})
\right)
+\alpha_\infty\left(
\frac{q^2}{2H\A}+{\cal O}(\A^{-2})
\right)   \nonumber \\
&+&C_\sigma\left(
\frac{u_\ast q^2 dP_1}{36(q^2+1) H\,\A^3}+{\cal O}(\A^{-4})
\right).
\eea
This gives the complete solution for perturbations around  \FFF background. 
The solution is a superposition of four modes proportional to the
four integration constants  $C_\sigma,\alpha_\infty,\beta_\infty,\phi_\infty$. 
Now, we remember that the background solution \eqref{BNVaa}
depends on three ``moduli parameters"  $q,\chi,t_0$. 
It is clear that the $\alpha_\infty$ mode describes simply the change of the 
background under the  shift $t_0\to t_0+\delta t_0$. Likewise, the $\phi_\infty$ mode describes 
the background change under the parameter variation $q\to q+\delta q$ while the $\beta_\infty$ 
mode is generated by the shift  $\chi\to \chi+\delta \chi$. Therefore, these three modes
are actually trivial and can be removed by fixing the background 
parameters.  As a result, the only 
non-trivial deformations of the background (within the ansatz under consideration) 
are described by the $C_\sigma$ mode. One has for such solutions at late times 
\be
\sigma \propto \A^{-4}\,,~~~~~\alpha \propto \beta \propto \phi \propto  \A^{-3}\,. 
\ee
Since all perturbations quickly vanish for $\A\to\infty$, it follows that  {\FFF solutions
are late time attractors.}

\section{Fully anisotropic solutions: formulation}
\setcounter{equation}{0} 

We now wish to construct fully anisotropic solutions 
described by Eqs.\eqref{22},\eqref{11}.

\subsection{Constraints  \label{VII}}

We note first of all that the 
second order equations \eqref{22} 
can be easily resolved with respect to $\ddot{A}$ and $\ddot{B}$. 
 However, it is not immediately obvious whether or not one can resolve the 
first order equations \eqref{11} with respect to  $\dot{F}$ and  $\dot{f}$. In fact, 
by investigating instead of Eqs.\eqref{11}
their differential consequences -- the conservation conditions \eqref{B} linear 
in the second derivatives $\ddot{F}$, $\ddot{f}$ -- 
one can show that this is impossible. 
Indeed, a closer inspection reveals that these equations cannot be resolved with respect to 
$\ddot{F}$, $\ddot{f}$ since the corresponding coefficient matrix is degenerate and 
for a particular linear combination of the two equations \eqref{B} the 
$\ddot{F}$ and $\ddot{f}$ terms drop out altogether. 
The implicit function theorem then tells us that the first order equations \eqref{11} 
cannot be resolved with respect to  $\dot{F}$ and  $\dot{f}$. We shall see this explicitly in the following analysis. 

Let us rewrite these two equations as 
\bea             \label{11a}
\xi=
\frac{F}{AY}\left(
\dot{F}+\frac{F}{A}
\right),  ~~~~~~~~~~~~
\zeta=\frac{F^2\dot{f}}{Y}\,,
\eea
where
\be
\xi\equiv -\frac{G^0_0+P_0}{P_1},~~~~~~
\zeta\equiv -\frac{G^0_x}{P_1},
\ee
with 
\bea             
G^0_0=\frac{3}{A^2}-\frac{2\dot{A}\dot{B}}{AB}-\frac{\dot{B}^2}{B^2},~~~~~~~~~~~
G^0_x=2\,\frac{\dot{B}}{B}-2\,\frac{\dot{A}}{A}. 
\eea
Using the definition of $Y$ in \eqref{b} it is not difficult to resolve each of the two equations \eqref{11a}
with respect to $\dot{F}$, which gives, respectively, two relations 
\bea               \label{F1}
\dot{F}+\frac{F}{A}&=& \frac{F\dot{f}}{\sqrt{1-\left({F}/{A\xi} \right)^2   }  }\,, \nonumber \\
\dot{F}+\frac{F}{A}&=& F\dot{f}\sqrt{1+\frac{F^2}{\zeta^2}}\,. 
\eea
As we have anticipated from the implicit function theorem, these do not determine 
both $\dot{F}$ and $\dot{f}$ since taking their ratio gives an algebraic relation not containing  $\dot{F},\dot{f}$ at all, 
\be
\frac{1}{\sqrt{1-\left({F}/{A\xi} \right)^2   }  }= \sqrt{1+\frac{F^2}{\zeta^2}}\,.
\ee
This implies that 
\be               \label{FFF}
F^2=A^2\xi^2-\zeta^2\,
\ee 
and also 
\bea                  \label{FFFF}
\dot{F}&=&-\frac{F}{A}+AF\,\frac{\xi}{\zeta}\,\dot{f}\,,\nonumber \\
Y&=&\frac{F^2\dot{f}}{\zeta}\,.
\eea
This  solves the first order Einstein equations \eqref{11}.  There remains to solve the
second order Einstein equations \eqref{22}. These 
contain in the right hand side terms with 
$F,\dot{F}$ which can be expressed by using \eqref{FFF}, \eqref{FFFF}.
Therefore, 
the $F$-amplitude can be eliminated from the problem altogether. 
However, the equations will still contain $f$ and $\dot{f}$, 
although we do not yet have 
an equation for the $f$-amplitude. 

To obtain the missing equation we rewrite \eqref{FFF}
in the form of constraint, 
\be               \label{F}
{\cal C}(A,B,\dot{A},\dot{B},u,F)=0,
\ee
where 
\be
{\cal C}= A^2 \left( \frac{3}{A^2}-\frac{2\dot{A}\dot{B}}{AB}-\frac{\dot{B}^2}{B^2}   
+P_0(u)\right)^2-4\left(\frac{\dot{B}}{B}-\frac{\dot{A}}{A}\right)^2-(P_1(u))^2 F^2=0. 
\ee 
This constraint should be preserved in time, hence
one should have 
\be
\dot{\cal C}=
\frac{\partial \cal C}{\partial A}\,\dot{A}+
\frac{\partial \cal C}{\partial \dot{A}}\,\ddot{A}+
\frac{\partial \cal C}{\partial B}\,\dot{B}+
\frac{\partial \cal C}{\partial \dot{B}}\,\ddot{B}+
\frac{\partial \cal C}{\partial F}\,\dot{F}+
\frac{\partial \cal C}{\partial {u}}\,\dot{u}=0.
\ee
Here $\ddot{A},\ddot{B}$ are determined by the Einstein  equations \eqref{22} while 
$\dot{F}$ is given by \eqref{FFFF} whereas the definition \eqref{uu}  of $u$ yields 
\be
\dot{u}=\left(\frac{\dot{F}}{F}-\frac{\dot{B}}{B}+\dot{f}\right)u.
\ee
As a result, $\dot{\cal C}$ is a function of $A,B,\dot{A},\dot{B},u,\dot{f}$. 
Explicitly, 
\be                            \label{dC}
\dot{\cal C}=2P_1(u)\left(A^2\frac{\xi}{\zeta}\,\dot{f}-1\right)(A\xi+\zeta)\,{\cal S}(A,B,\dot{A},\dot{B},u)\,,
\ee
where 
\bea
{\cal S}(A,B,\dot{A},\dot{B},u)&\equiv &\left(
P_0\,A^2 B^2 -A^2 \dot{B}^2-2AB\dot{A}\dot{B}-2AB\dot{B}+2B^2\dot{A}+3B^2
\right) \times \nonumber \\
&\times &\left[
\frac{P_0\, A^3 B^2 \dot{B}-A^3 \dot{B}^3 -2A^2 B\dot{A}\dot{B}^2 +AB^2 \dot{B}+2B^3 \dot{A}}{A^4 B^5}\,dP_1
\right.  \\
&+&\left.\frac{u(A\dot{B}+B)}{A^2 B^3}\, P_1 dP_1-\frac{2A\dot{B}+B\dot{A}-B}{A^2 B^3} \, P_1^2
\right]+u\,(u\,dP_1-2P_1)P_1^2.   \nonumber 
\eea
It follows that the condition $\dot{\cal C}=0$ can be fulfilled in three different ways\footnote{
Setting $P_1(u)=0$ in \eqref{dC} would bring us back to the isotropic case.}. 
First, one could set 
$
\dot{f}={\zeta}/{A^2\xi}
$
which would give  the missing equation for $f(t)$, but 
Eq.\eqref{FFFF} would then  yield $\dot{F}=0$, hence the reference metric \eqref{BNVf}
would be degenerate. Therefore, this option is not interesting. 
Secondly, one could set $A\xi+\zeta=0$, but Eq.\eqref{FFF} would then yield $F=0$, hence this option 
is not interesting either. Therefore, the third factor in \eqref{dC} must vanish, i.e. 
${\cal S}(A,B,\dot{A},\dot{B},u)=0$. This is the secondary constraint that insures 
the stability of the primary constraint ${\cal C}=0$. 
Now, the secondary constraint must be stable as well, hence one should have 
\be
\dot{\cal S}=
\frac{\partial \cal S}{\partial A}\,\dot{A}+
\frac{\partial \cal S}{\partial \dot{A}}\,\ddot{A}+
\frac{\partial \cal S}{\partial B}\,\dot{B}+
\frac{\partial \cal S}{\partial \dot{B}}\,\ddot{B}+
\frac{\partial \cal S}{\partial {u}}\,\dot{u}=0.
\ee
A straightforward (but lengthy) calculation shows that 
\be
\dot{\cal S}={\cal W}(A,B,\dot{A},\dot{B},u)\,\dot{f}+{\cal V}(A,B,\dot{A},\dot{B},u),
\ee
where ${\cal W,V}$ are rather complicated functions that we do not write down. 
Therefore,  setting $\dot{\cal S}=0$ does not give a tertiary constraint but rather the 
condition that determines $\dot{f}$, 
\be                                \label{f1}
\dot{f}=-\frac{{\cal V}(A,B,\dot{A},\dot{B},u)}{{\cal W}(A,B,\dot{A},\dot{B},u)}\equiv {\cal F}(A,B,\dot{A},\dot{B},u).
\ee
This is the missing equation. 

\subsection{Equations \label{VIII}}

Summarising the above discussion, 
the two constraints ${\cal C}=0$ and ${\cal S}=0$ allow us to algebraically 
express the Stuckelberg fields $F$ and $f$ and their first derivatives in terms of $A,B,\dot{A},\dot{B}$. 
As a result, the problem reduces to integrating the second order equations for $A$ and $B$. 

It is, however, convenient to consider the  Stuckelberg fields as dynamical variables alongside with $A$ and $B$ 
and impose the constraints only at the initial time moment. We choose the independent variables to be 
$A,B,F,u$.  The corresponding equations are 
\bea                  \label{eeqs}
\dot{F}&=&F\left[-\frac{1}{A}+A\,\frac{\xii}{\psii}\,{\cal F}\right], \nonumber \\
\dot{u}&=&u\left[-\frac{1}{A}-\frac{\dot{B}}{B}+\left(A\,\frac{\xii}{\psii}+1\right){\cal F}\right], \nonumber \\
\ddot{B}&=&B\left[-\frac{\dot{B}^2}{B^2}-\frac{\dot{A}\dot{B}}{AB}+\frac{2}{A^2}
+P_0(u)+\frac12\,Y P_1(u)\right],   \nonumber   \\
\ddot{A}&=&A\left[-\frac{2\dot{A}\dot{B}}{AB}+\frac{2}{A^2}
+P_0(u) -\left(
u-\frac12\,Y-\frac{F^2}{Y}\frac{\xii}{\psii}\,{\cal F} \right)P_1(u)  \right. \nonumber \\
&&~~~~~~-\left.\frac12\left(
Yu-u^2
+\frac{F^2}{A^2}-F^2\frac{\xii}{\psii}\,{\cal F}
\right)dP_1(u)\right], 
\eea
where 
\be            
\xii=-\frac{1}{P_1(u)}\left(\frac{3}{A^2}-\frac{2\dot{A}\dot{B}}{AB}-\frac{\dot{B}^2}{B^2}+P_0(u)\right),~~~~
\psii=-\frac{2}{P_1(u)}\left(\frac{\dot{B}}{B}-\frac{\dot{A}}{A}\right),~~~~
Y=\frac{F^2{\cal F}}{\psii}\,,
\ee
while 
\be
{\cal F}={\cal F}(A,B,\dot{A},\dot{B},u)
\ee
is defined by \eqref{f1}. To start the integration one chooses initial values $A_0,B_0,\dot{A}_0,\dot{B}_0$
and solves 
the secondary constraint 
${\cal S}(A_0,B_0,\dot{A}_0,\dot{B}_0,u_0)=0$ to obtain 
\be           \label{uuu}
u_0=u_0(A_0,B_0,\dot{A}_0,\dot{B}_0).
\ee
Then one solves  the primary constraint 
${\cal C}(A_0,B_0,\dot{A}_0,\dot{B}_0,u_0,F_0)=0$ to obtain 
\be
F_0=F_0(A_0,B_0,\dot{A}_0,\dot{B}_0). 
\ee
This gives initial values for equations \eqref{eeqs}. Integrating the equations, 
the constraints should be preserved in time, which gives a good consistency check. 

Let us finally comment on the sign choice. The f-metric \eqref{BNVf}, the ${\cal C}$-constraint \eqref{F}, 
and the equations \eqref{eeqs} are invariant under $F\to-F$, hence $F$ is defined only up to a sign,
but since $F=uBe^{-f}$, its sign should be   chosen the same as that of $u$. The latter is determined 
unambiguously, since the initial value of $u$ is determined by the ${\cal S}$-constraint, which is not 
invariant under $u\to -u$. 

\section{Numerical results \label{IX}}
\setcounter{equation}{0}

A comprehensive  analysis of solutions of equations \eqref{eeqs} is a difficult task. 
The equations contain four parameters $b_k$  and four other 
parameters $A_0,B_0,\dot{A}_0,\dot{B}_0$ determine the initial data, 
hence the space of solutions is eight dimensional. 
In addition, for given values of the eight parameters there can be 
several solutions of the constraint
${\cal S}(A_0,B_0,\dot{A}_0,\dot{B}_0,u_0)=0$
 determining  the initial value $u_0$. 
 As a result,  there can be many  different solutions. 
 Nevertheless, we were able to identify just three basic solution types. 
 They are obtained either  for random initial values, or for initial values corresponding to 
perturbed \FF solution.  Maybe there exist also some other 
 solution types, but we have not been able to  detect them. 
  
 \begin{figure}[th]
\hbox to \linewidth{ \hss
	\resizebox{8cm}{7cm}{\includegraphics{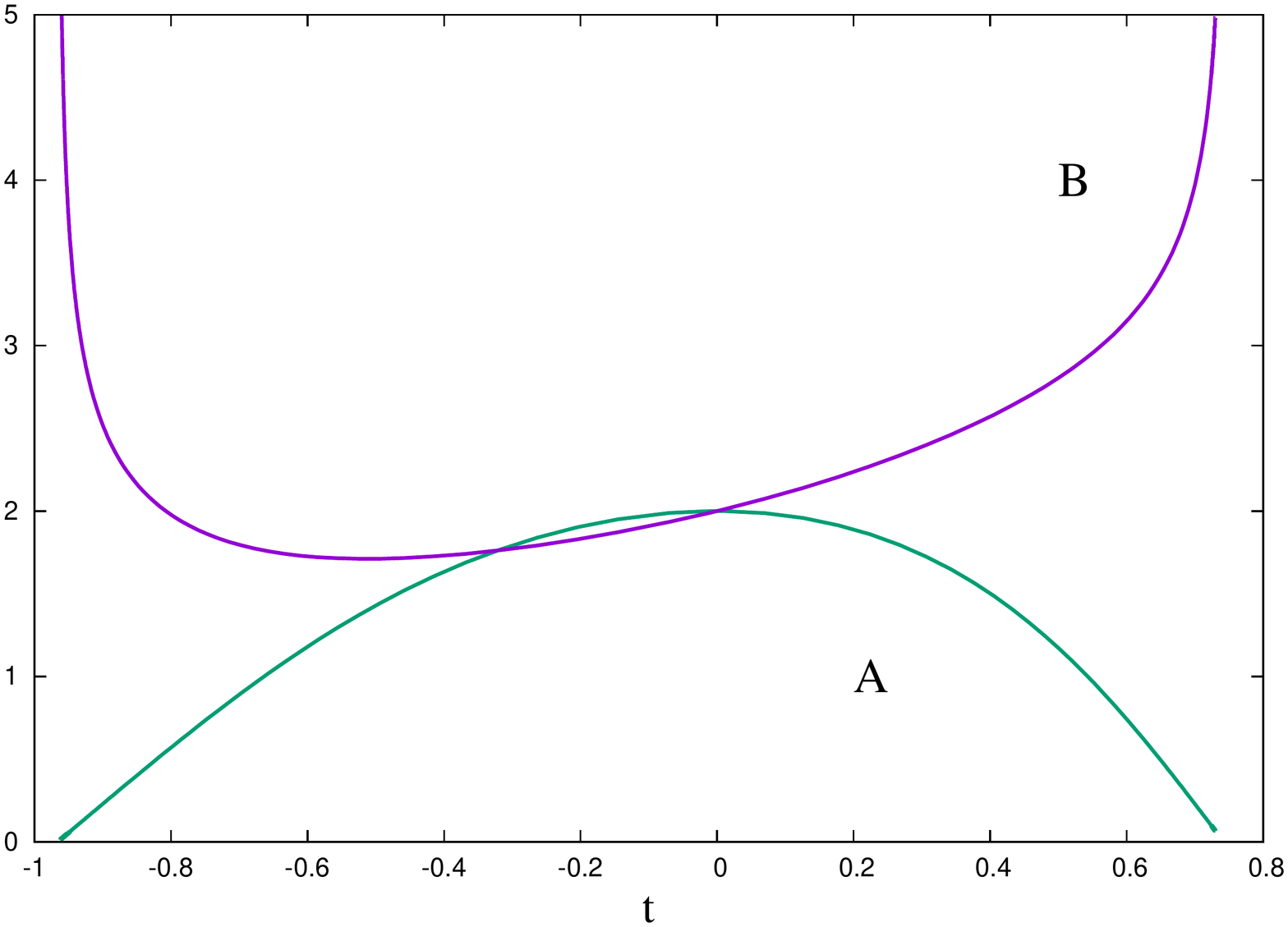}}
\hspace{1mm}
	\resizebox{8cm}{7cm}{\includegraphics{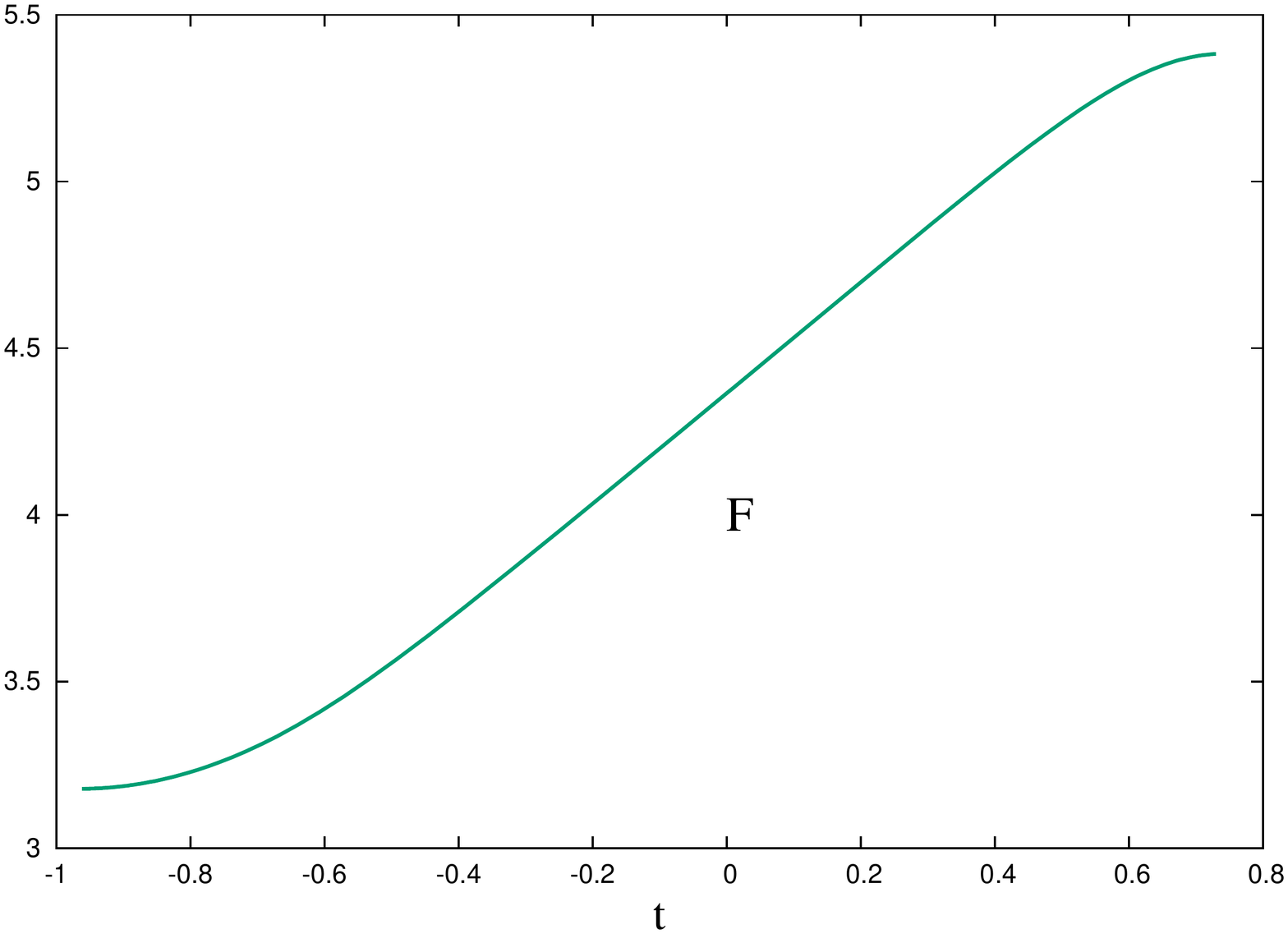}}
\hspace{1mm}
\hss}
\hbox to \linewidth{ \hss
	\resizebox{8cm}{7cm}{\includegraphics{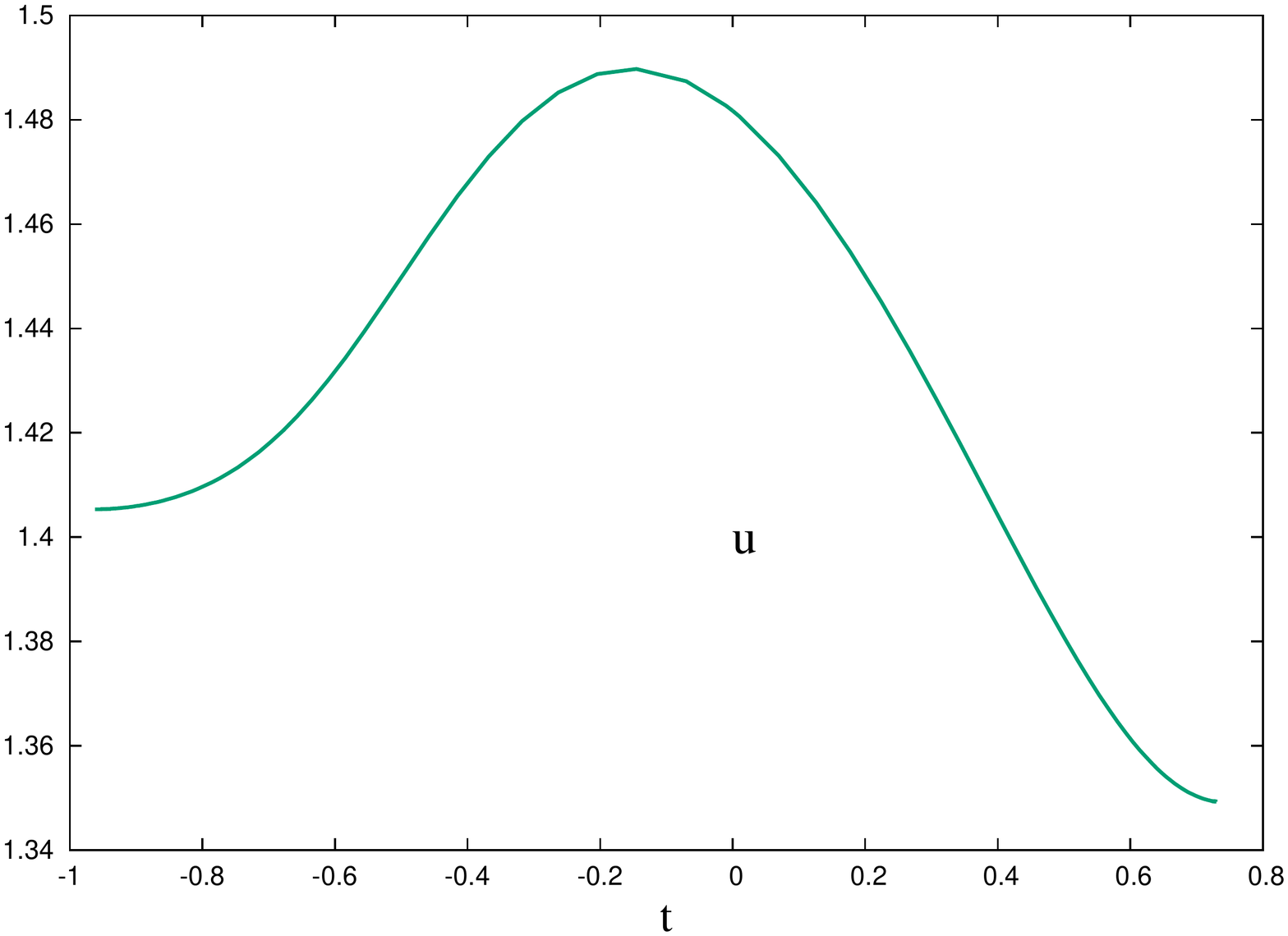}}
\hspace{1mm}
	\resizebox{8cm}{7cm}{\includegraphics{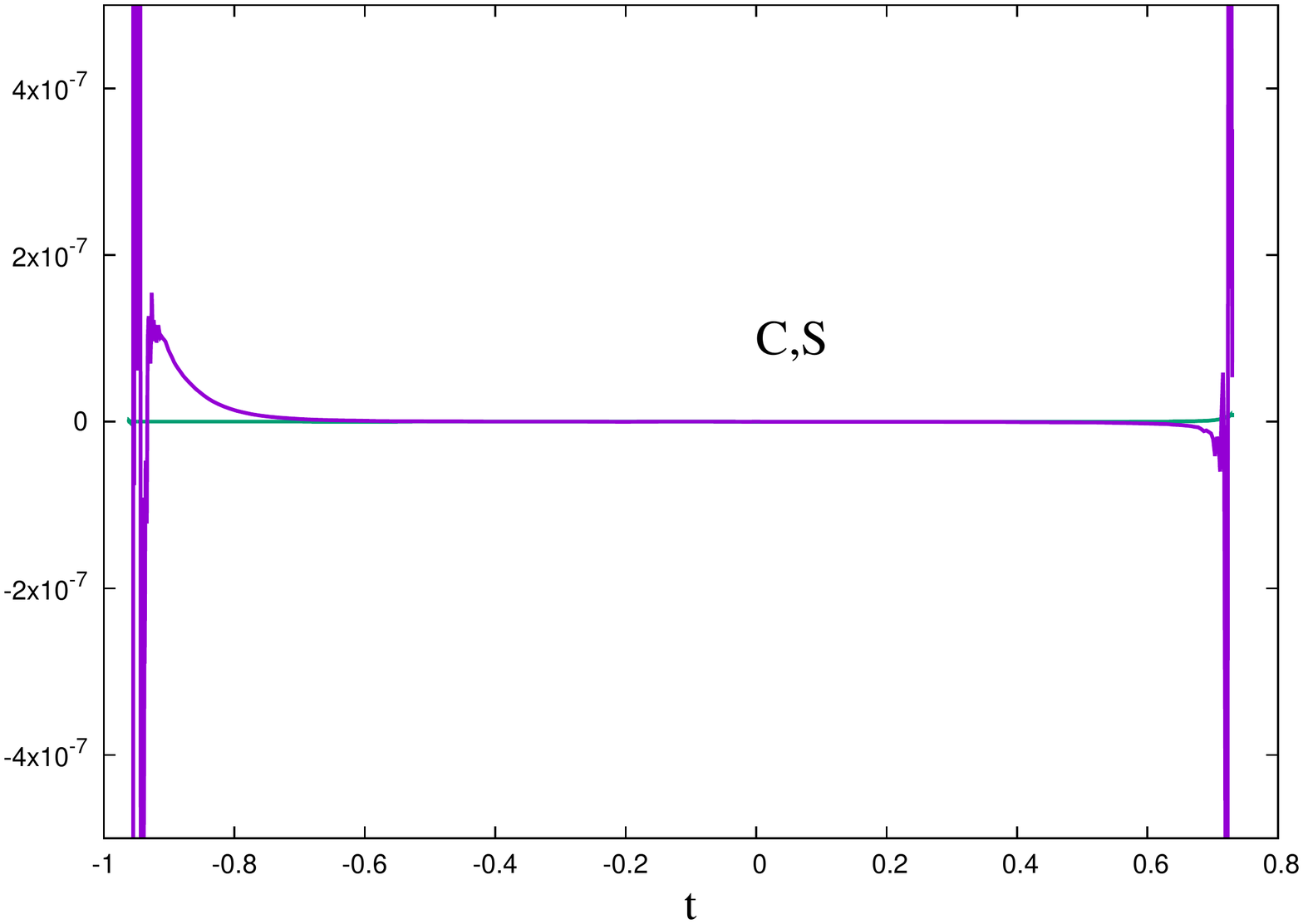}}
\hspace{1mm}
\hss}
\caption{Evolution of initial data \eqref{in1}--\eqref{in3}: singularity formation. }
\label{Fig}
\end{figure}

\subsection{Generic initial values} 

Let us choose some values for the theory parameters, for example 
\be                      \label{bbb}
b_0=1,~~~b_1=1,~~~b_2=2,~~~b_3=-5. 
\ee
We choose next some arbitrary initial values for which the universe 
is anisotropic already at the initial time moment $t=0$. One should emphasise that the ``initial moment''
has nothing to do with the initial singularity but simply labels  the 
timelike hypersurface containing the Cauchy data. 
For example, we chose 
\be                     \label{in1}
A_0=B_0=2,~~~~~\dot{A}_0=0,~~~~\dot{B}_0=1.
\ee
The equation ${\cal S}(u_0)=0$ then shows two real roots, one of which is 
\be                      \label{in2}
u_0=1.4817. 
\ee
Using this, the equation ${\cal C}(F_0)=0$ gives 
\be               \label{in3}
F_0=4.3649.
\ee
Integrating the equations with these initial conditions starting from $t=0$ towards $t>0$ 
and then towards $t<0$ gives the result shown in Fig \ref{Fig}.

The numerical solution extends over a finite interval. Close to its ends
 the $A$ amplitude becomes small and visibly approaches zero while the derivative $\dot{B}$ grows. 
This suggests that at the ends of the interval $A$ vanishes and 
 the solution develops a curvature singularity which is difficult to  approach numerically.
 At the same time,  nothing visibly special happens to the $F$ and $u$ amplitudes. The constraints 
 ${\cal C}$ and ${\cal S}$ 
  both remain 
of the order of $10^{-9}$ and  start to grow only close to the ends of the interval. 
Changing values of  $b_k$ and $A_0,B_0,\dot{A}_0,\dot{B}_0$
we find that 
this type of behaviour is typical -- generic  solutions  develop 
singularities where one of the metric amplitudes vanishes and/or derivatives 
of other fields amplitudes grow. 
To avoid such a singular behaviour,  we fine-tune the initial values.

\subsection{ Slightly perturbed \FF}

Let us see what happens if the initial values are close to  \FF solution. 
Choosing again the parameters $b_k$ according to  \eqref{bbb}, 
the equation $P_1(u_\ast)=0$ has two roots: 
\bea             \label{pars}
u_\ast&=&-\frac15,~~~~~~~P_0(u_\ast)=0.68,~~~~~~~H(u_\ast)=\sqrt{P_0(u_\ast)/3}={0.476};\nonumber \\
u_\ast&=&1,~~~~~~~~~~P_0(u_\ast)=5,~~~~~~~~~~~H(u_\ast)=\sqrt{P_0(u_\ast)/3}=1.29.
\eea
Since for each of these roots one has $P_0(u_\ast)>0$ 
(which is not the case for arbitrary  values of $b_k$), the cosmological constant $P_0(u_\ast)/3$ is positive,
hence  each root gives rise to a \FF solution with its own Hubble rate $H(u_\ast)$.

Let us select the first root in \eqref{pars}, $u_\ast=-1/5$, 
and then 
choose the initial values of $A,B,\dot{A},\dot{B}$ to be 
``almost" \FF (we set here $\A=10$), 
\be                          \label{data}
A_0=B_0=\A,~~~~~\dot{A}_0=\sqrt{1+H^2(u_\ast)\A^2},~~~~\dot{B}_0=\dot{A}+\delta. 
\ee
For $\delta=0$ these values are precisely \FFd
To make them ``slightly anisotropic'' we choose 
$\delta=-0.1$. Then the initial value $u_0$ is no longer exactly $u_\ast=-0.2$
but is determined by the ${\cal S}(u_0)=0$ constraint, which has  four real roots,
\be                     \label{root}
u_0^{(1)}=-0.231122,~~~~u_0^{(2)}=-0.233943,~~~~u_0^{(3)}=-0.152569,~~~~u_0^{(4)}=-0.645204.
\ee
The ${\cal C}(F_0)=0$ constraint then gives, correspondingly,  the values 
\be                    
F_0^{(1)}=-0.831254,~~~~F_0^{(2)}=-0.905497,~~~~F_0^{(3)}=-2.95427,~~~~F_0^{(4)}=-0.323448. 
\ee
\begin{figure}[th]
\hbox to \linewidth{ \hss
	\resizebox{8cm}{7cm}{\includegraphics{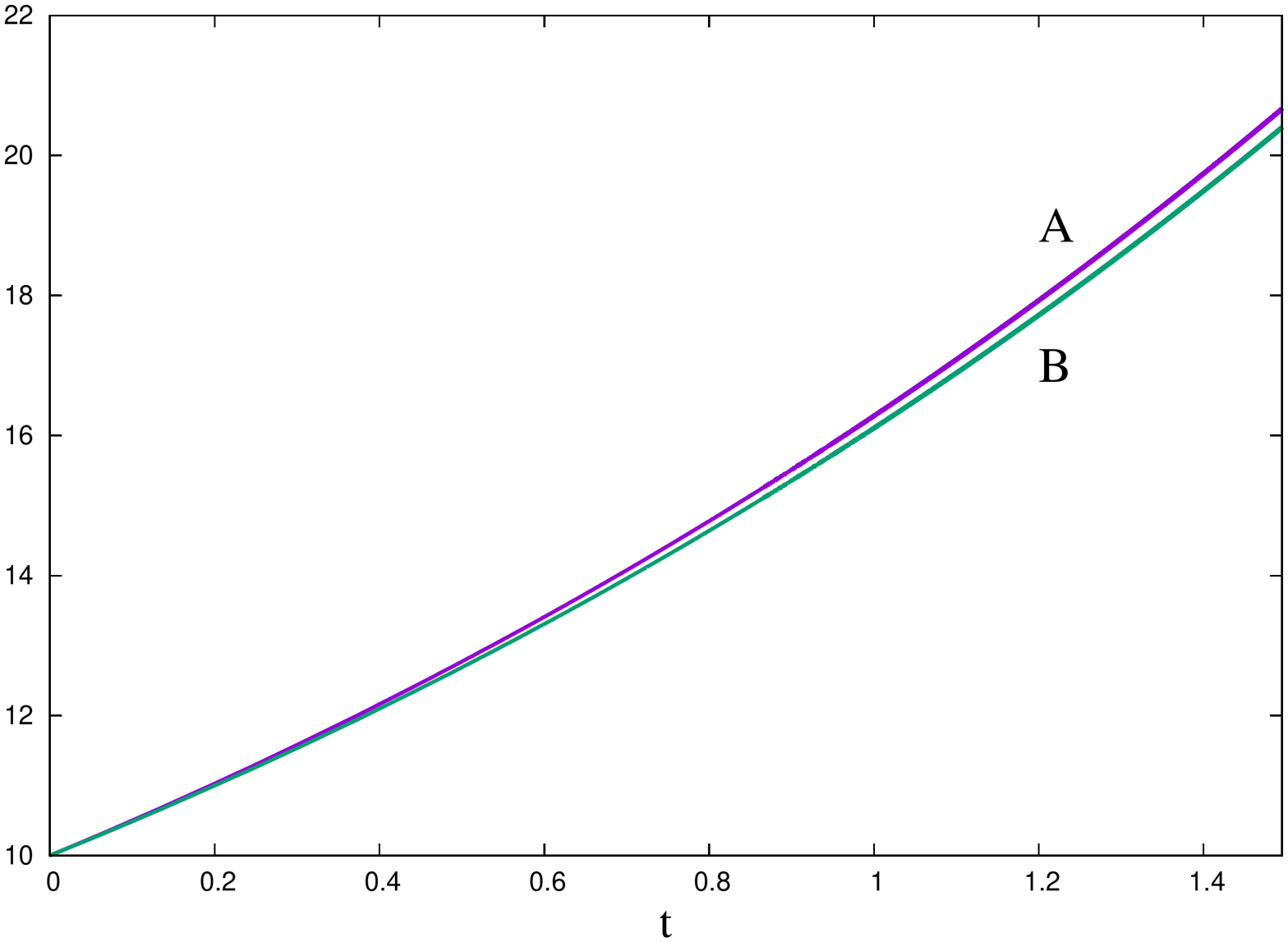}}
\hspace{1mm}	
	\resizebox{8cm}{7cm}{\includegraphics{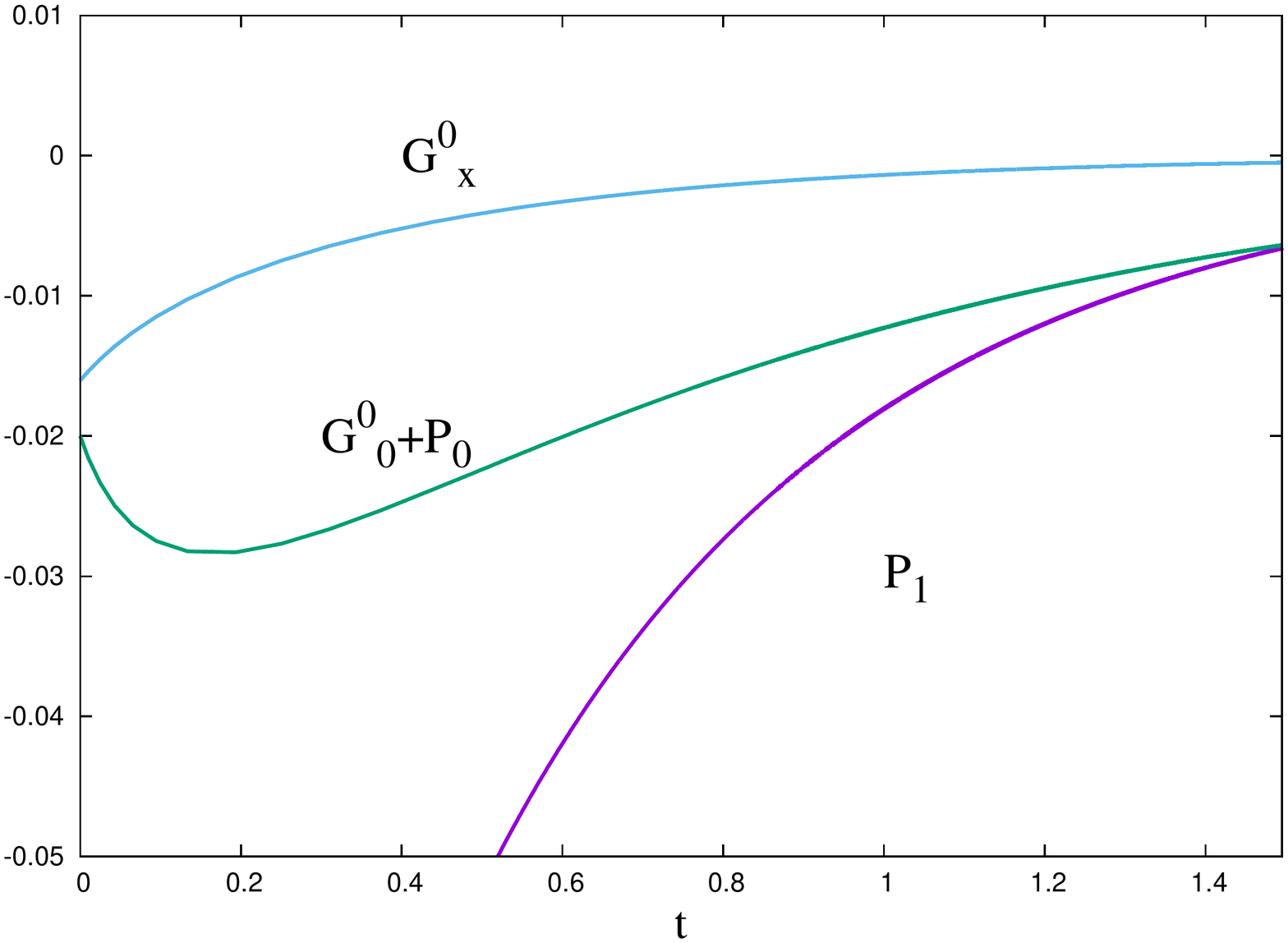}}	
\hspace{1mm}
\hss}
\hbox to \linewidth{ \hss
	\resizebox{8cm}{7cm}{\includegraphics{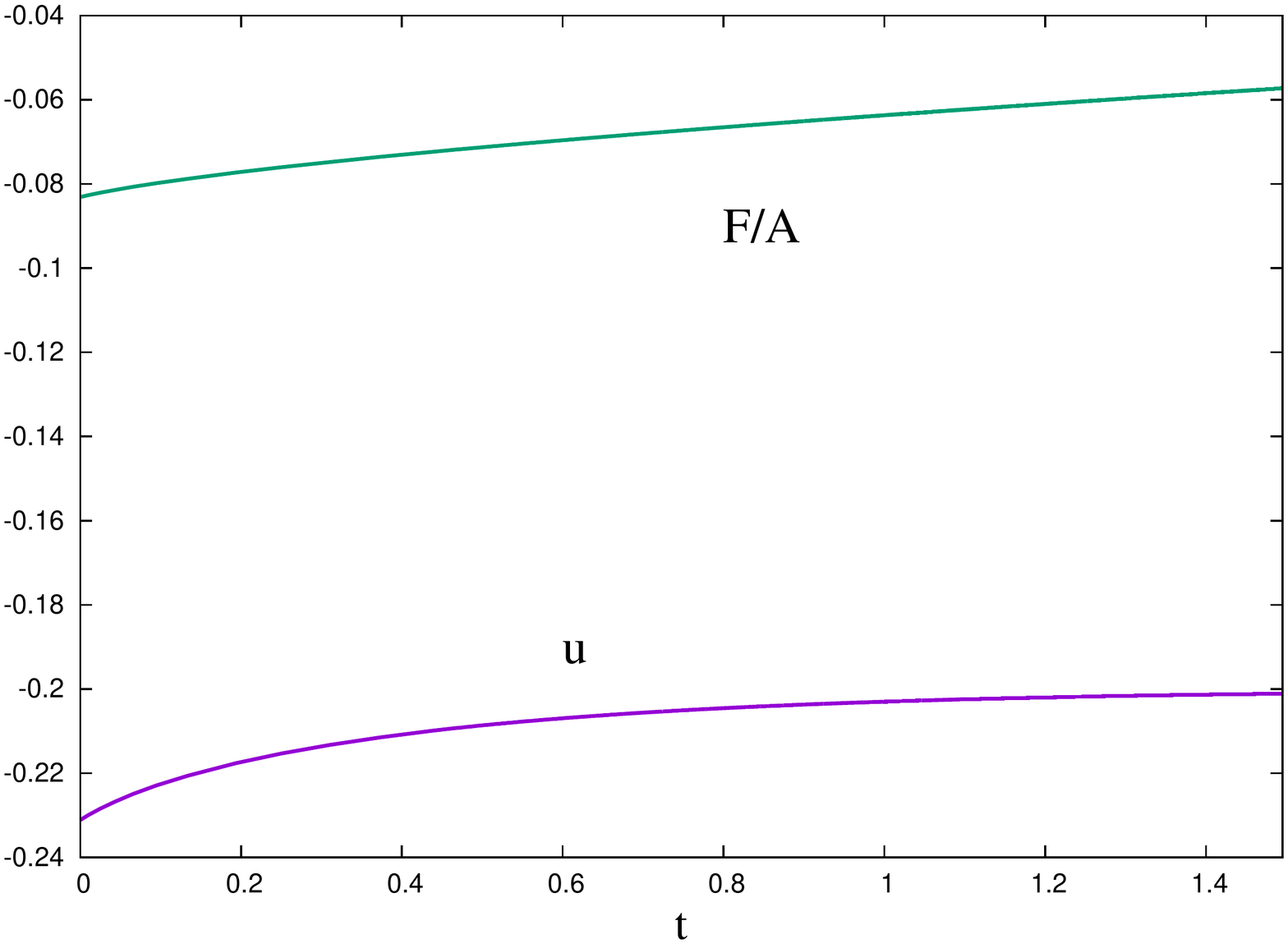}}
\hspace{1mm}	
	\resizebox{8cm}{7cm}{\includegraphics{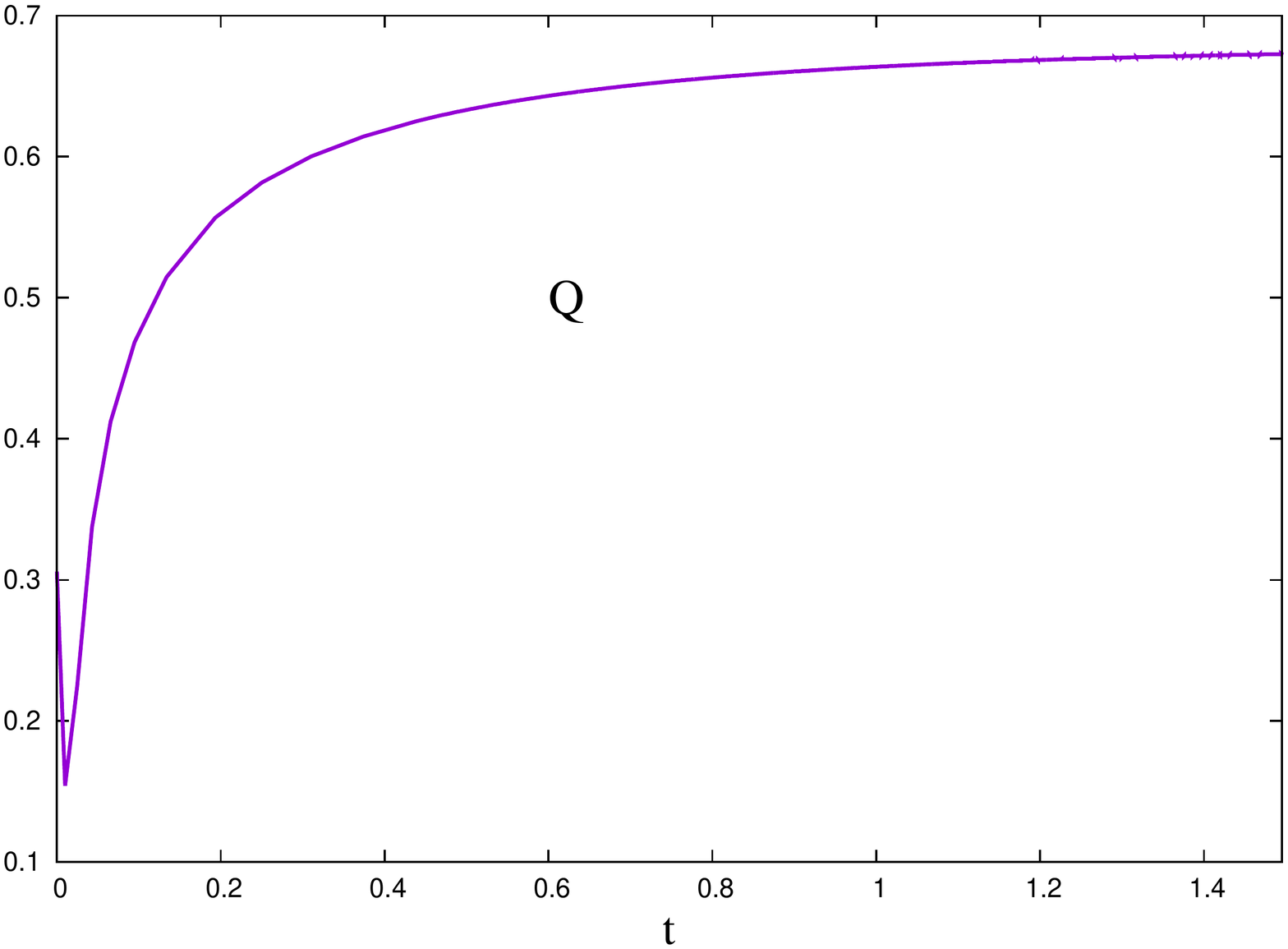}}	
\hspace{1mm}
\hss}
\caption{Evolution of a weakly perturbed type I solution  -- relaxation to \FFFd}
\label{Fig2}
\end{figure}
We see that the values $u_0^{(1)}$ and $u_0^{(2)}$ are closer to $u_\ast=-0.2$  
than $u_0^{(3)}$ and $u_0^{(4)}$. Therefore, although the g-metric is almost isotropic, 
the Stueckelberg fields in the  latter two cases are far  from  \FF value, hence such 
initial values actually corresponds to a {\it strong} perturbation. 
This is 
confirmed by the numerics -- solutions generated by the initial choice 
$u_0=u_0^{(3)}$ or $u_0=u_0^{(4)}$ develop a curvature  singularity 
similar to that discussed in the previous subsection. 

Let us now see what happens if $u_0=u_0^{(1)}$ or $u_0=u_0^{(2)}$
so that the initial values are closer to   \FF configuration. 
It turns out that solutions obtained in these two cases are almost identical 
and we therefore describe only the $u_0=u_0^{(1)}$ solution shown in Fig.\ref{Fig2}.  

As one can see  in Fig.\ref{Fig2}, the $A$ and $B$ amplitudes always stay very close to each other,
while the whole configuration becomes ``more and more isotropic''. 
Indeed, both for type I and \FFF isotropic solutions one has $P_1=G^0_x=G^0_0+P_0=0$ and $u=u_\ast$.
At the same time,  one sees in Fig.\ref{Fig2} that $P_1$, $G^0_x$ and $G^0_0+P_0$ approach zero while
$u$ approaches $u_\ast$. Therefore, the solution approaches either type I or \FFFd  Now, 
if it was \FF then the ratio $F/A$ would approach $u_\ast$, which is clearly 
not the case as is seen in Fig.\ref{Fig2}. Therefore, the solution must approach \FFFd
To verify this we plot in Fig.\ref{Fig2} the function 
\be
Q=\sqrt{\frac{F\dot{F}}{u_\ast^2 A}}.
\ee
For \FFF solutions \eqref{ww2} this functions assumes a constant value $Q(t)=q$, which  is the 
integration constant in \eqref{ww2}. For our solution, as is seen in Fig.\ref{Fig2}, $Q(t)$ approaches 
a constant value, hence the solution indeed approaches the isotropic \FFF background \eqref{ww2}
with $q=Q(\infty)$. 

We find a similar behaviour also for all other choices  of the theory parameters $b_k$ that we considered. 
It is difficult to extend numerical solutions to large values of $A,B$ since the constraints start to grow,
but using the multi-shooting method we managed to keep them under control and 
extend the solutions  to the region 
where  $P_1$, $G^0_x$ and $G^0_0+P_0$ become very small
while $Q(t)$ become almost constant. 
Since such solutions seem to exist for generic parameter values, we conclude that slightly 
perturbed type I configurations evolve towards \FFF isotropic fixed point \eqref{ww2}.

\begin{figure}[th]
\hbox to \linewidth{ \hss
	\resizebox{8cm}{7cm}{\includegraphics{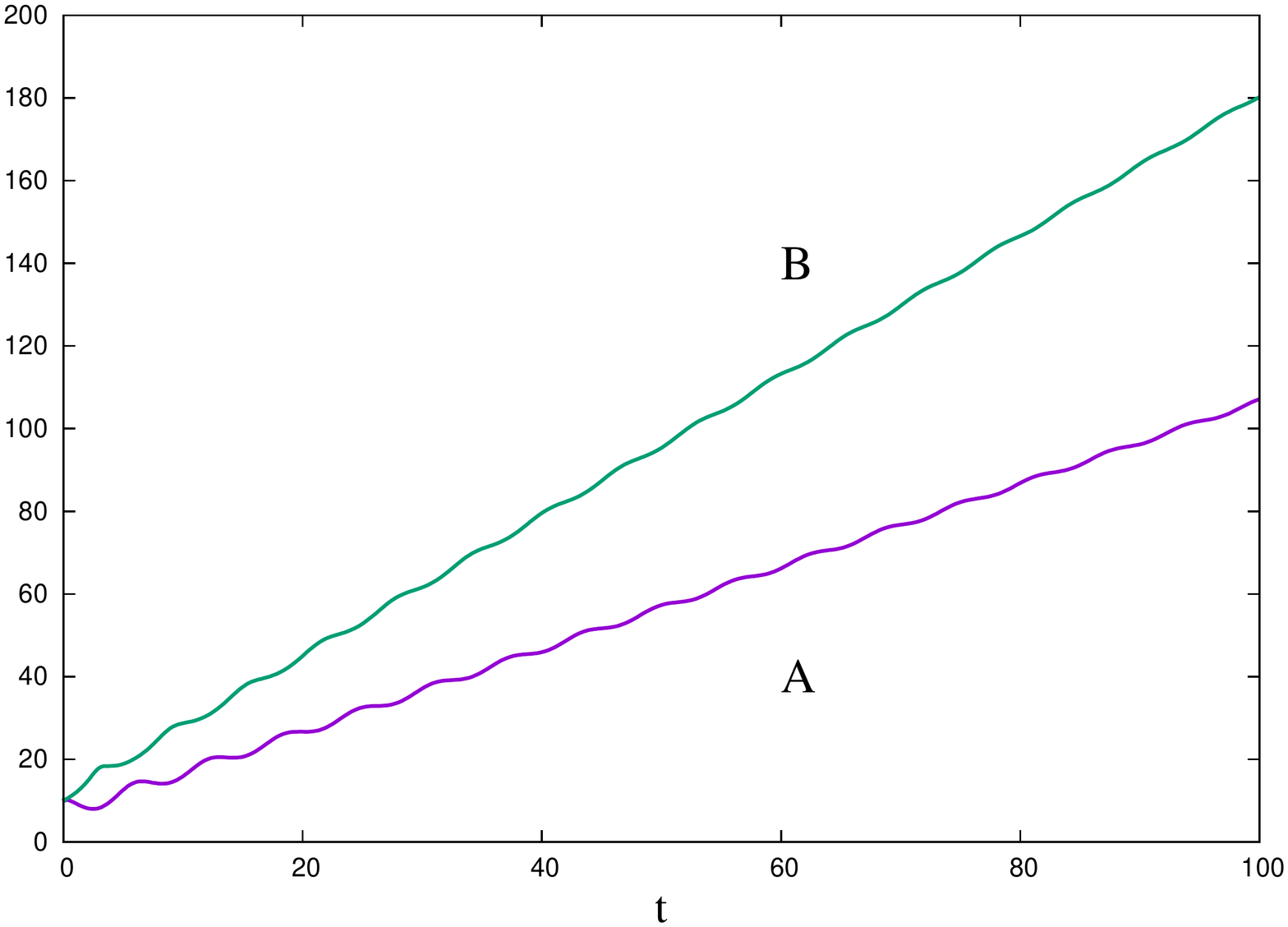}}
\hspace{1mm}	
	\resizebox{8cm}{7cm}{\includegraphics{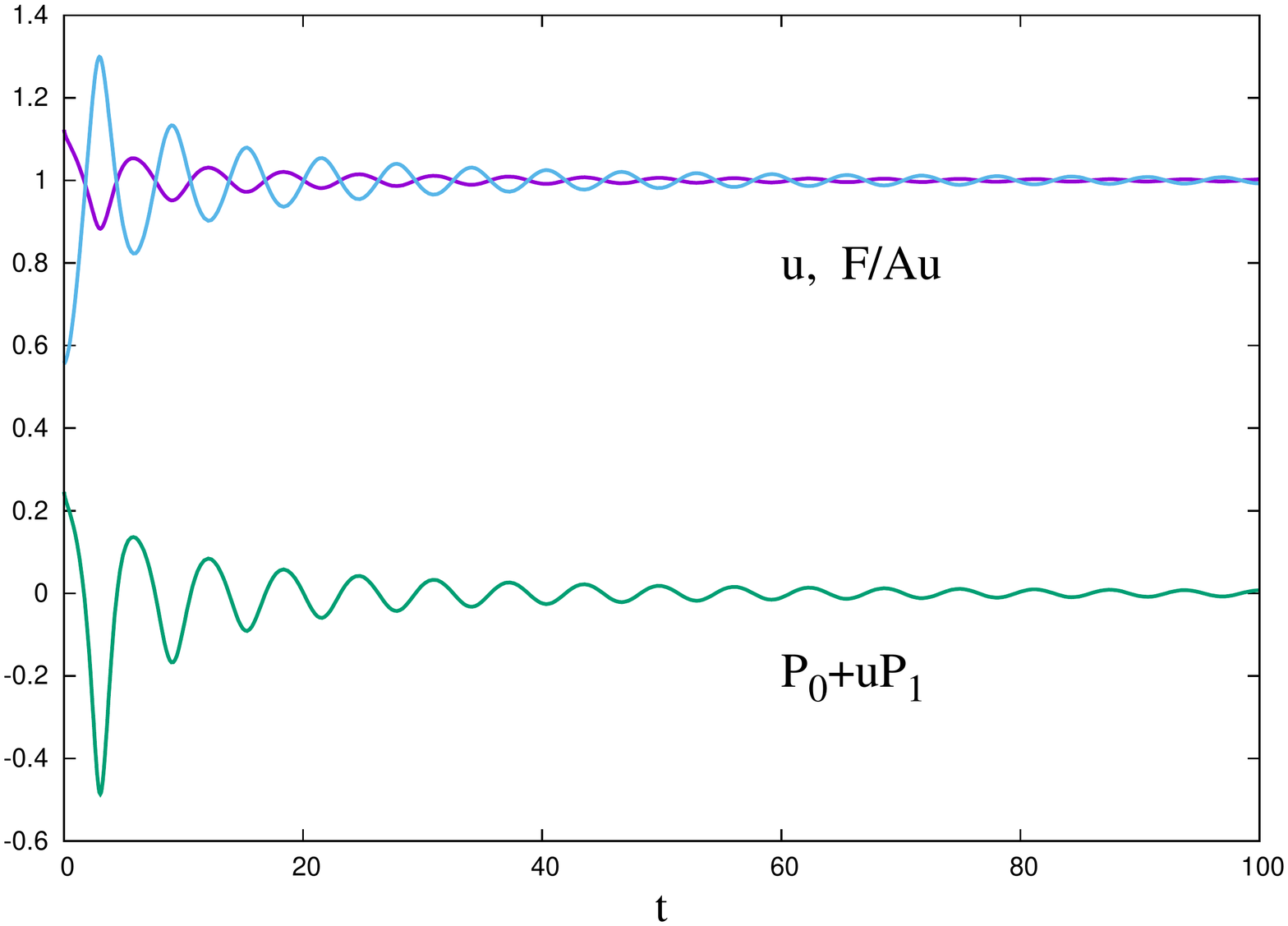}}	
\hspace{1mm}
\hss}
\hbox to \linewidth{ \hss
\hspace{1mm}	
\hspace{1mm}
\hss}
\caption{Evolution of strongly perturbed \FF  -- decay into flat space. }
\label{Fig3}
\end{figure}

\subsection{ Strongly perturbed \FF -- decay into flat space}

We have already mentioned above what is meant by strong perturbations -- parameterising 
the initial values similarly to \eqref{data}  and choosing the root of 
${\cal S}(u_0)=0$ to be far from the root of $P_1(u_\ast)=0$. As a result, the physical geometry is 
initially close to that for  \FF  solution but the Stueckelberg fields 
are different. As was mentioned above,  the evolution of such initial data 
generically leads to a curvature singularity.  However, we were able to find  parameter values for which 
the outcome is different. Specifically, choosing 
\be
b_0= -19,~~~~
b_1= 14, ~~~~
b_2 = -10,~~~~
b_3= 7,
\ee
one root of  $P_1(u_\ast)=0$ is $u_\ast=1.63$ with $H(u_\ast)=\sqrt{P_0(u_\ast)/3}=0.15$.
Using this to compute 
$A_0,B_0,\dot{A}_0,\dot{B}_0$ in \eqref{data} (with $\A=10$ and $\delta=-0.1$) 
and then solving the ${\cal S}(u_0)=0$ constraint gives four real roots, 
\be
u_0^{(1)}=1.1222,~~~~~
u_0^{(2)}=1.5909,~~~~~
u_0^{(3)}=1.6362,~~~~~
u_0^{(4)}=1.6680.
\ee 
The root $u_0^{(3)}=1.6362$ is the closest to $u_\ast=1.63$  and gives rise to a
slightly perturbed type I configuration  that relaxes to \FFFd 
Let us consider   instead  $u_0^{(1)}=1.1222$ --  the 
{\it farthest} from $u_\ast=1.63$ root. Surprisingly, the evolution of this initial data does not 
lead to a singularity but to something different -- a decay into flat spacetime. 
As shown in Fig.\ref{Fig3}, the fields show damped oscillations and at late times
the $A$ and $B$ amplitudes become 
linear functions of time, $u$ approaches a constant value such that the combination  $P_0+uP_1$ 
tends to zero, while ${\cal A}=F/Au$ tends to one. Therefore, the fields approach 
the flat spacetime solution described by Eqs.\eqref{A55}--\eqref{A9} in Appendix~\ref{AppA}:
\be
A=\A=t,~~~~~B=Ae^\chi,~~~~F={\cal A}\A u,~~~~{\cal A}=1,~~~~~P_0(u)+uP_1(u)=0. 
\ee
It should be emphasised that we did not find such solutions for generic  $b_k$.

\section{Conclusions \label{X}}

To recapitulate, we studied above the fully  non-linear dynamics of anisotropic deformations 
of the homogeneous and isotropic cosmology in the ghost free massive gravity with flat reference metric. 
We found that when perturbed, this solution cannot relax to itself in the long run, hence it is unstable. 
If the initial perturbation is not too strong, it relaxes instead to \FFF solution whose 
physical metric is also de Sitter. Therefore, the  geometry described by the physical g-metric 
is stable and does relax to itself. 
However, during the relaxation and damping of the anisotropies 
the Stueckelberg scalars change  in such a way that 
the f-metric evolves 
from type I to \FFF value and looses some of the isometries that were common for both metrics.

The final \FFF configuration seems to be an attractor within the considered class of  anisotropic metrics.
This is confirmed by the analysis of linear modes in its vicinity and also by the numerics which show that 
slightly perturbed \FF configurations evolve towards \FFF solutions. It is natural 
to wonder if the latter is itself stable with respect to more general deformations. We leave this issue 
as well as the problem of detecting  possible ghosts to a separate study. 

If the initial perturbation is strong, then the initially isotropic solution completely changes its 
structure.  In the generic case it ends up in a singular state, but for some parameter values it can also 
decay into flat space. To pin down the parameter regions where the latter possibility is realised 
requires a separate study.

\section*{Acknowledgements}
S.M. acknowledges warm hospitality at the LMPT in Tours, where this work was initiated. 
His work was supported by Japan Society for
the Promotion of Science (JSPS) Grants-in-Aid for Scientific Research
(KAKENHI) No. 24540256, and by World Premier International Research
Center Initiative (WPI), MEXT, Japan.

M.S.V. thanks for hospitality the YITP in Kyoto, where this text  was completed. 
His work  was also partly supported by the Russian Government Program of Competitive Growth 
of the Kazan Federal University.

\appendix
\setcounter{section}{0}
\setcounter{equation}{0}
\setcounter{subsection}{0}
\section{Isotropic solutions with either $\dot{f}=0$ or $F=0$ \label{AppA}}

\renewcommand{\theequation}{\Alph{section}.\arabic{equation}}

We describe in this Appendix the remaining solutions of equations \eqref{222},\eqref{111}. 
Specifically, to solve the second equation in \eqref{111}, 
\be            \label{A0}
\frac{F^2}{Y}\,\dot{f}P_1=0,
\ee
it was assumed in Section \ref{isot} that $P_1=0$.
Let us now consider the other two options and assume  first that 
$
\dot{f}=0.
$
 Then  \eqref{b} 
 implies that 
$
Y=\dot{F}+{F}/{\A},
$
in which case  equations \eqref{222} and the first equation in \eqref{111} reduce to 
\bea                  
E_0&\equiv&\frac{3\dot{\A}^2}{\A^2}-\frac{3}{\A^2}=P_0
+\frac{F}{\A }P_1,  \label{AA1}  \\
E_1&\equiv&\frac{\ddot{\A}}{\A}+\frac{2\dot{\A}^2}{\A^2}-\frac{2}{\A^2}
=P_0+\frac12\,\left(\dot{F}+\frac{F}{\A}\right) P_1\,,   \label{AA2}  \\
0
&=&
\left(u-\frac{F}{\A}\right)
\left(P_1+\frac12(\dot{F}-u)dP_1\right).
 \label{AA3}  
\eea 
Due to the Bianchi identity,
\be
\dot{E}_0=6\,\frac{\dot{a}}{a}\,(E_1-E_0),
\ee
equation \eqref{AA2} can be replaced by
\be                        \label{A5}
\left(P_0+\frac{F}{\A}\,P_1\right)^{\mbox{.}}=3\,\frac{\dot{\A}}{\A}\,\left(\dot{F}-\frac{F}{\A}\right)P_1\,.
\ee
Next, \eqref{uu} and \eqref{prop} imply 
\be              \label{A55}
F=e^{\chi-f}\A u\equiv \cc\A u\,,
\ee
with constant $\cc$. Inserting this to \eqref{AA1},\eqref{AA3},\eqref{A5}  one obtains 
\bea                        \label{A6}
\left(P_0+\cc uP_1\right)^{\mbox{.}}&=&3\cc\,\frac{\dot{\A}}{\A}\,\left( (\A u)^{\mbox{.}}-u\right)P_1,   \\
(1-\cc) &u& \left(2P_1+[\cc(\A u)^{\mbox{.}}-u]dP_1\right)=0,   \label{A7} \\
\frac{3}{\A^2}(\dot{\A}^2-1)&=&P_0+\cc u P_1.    \label{A8}
\eea
Equation \eqref{A7} contains the product of three factors.

Let us assume that $\cc=1$, hence the first factor in \eqref{A7} vanishes and  the equation is fulfilled. 
Using $\dot{P}_m=dP_m\,\dot{u}$
and the relation \eqref{iden}, equation \eqref{A6} then reduces to 
$$
\dot{u}=\frac{\dot{\A}}{\A}\,\left( (\A u)^{\mbox{.}}-u\right),
$$
which is equivalent to 
\be                               \label{A9}
(\A u)^{\mbox{.}}(\dot{\A}-1)=0.
\ee
One solution of this is $\dot{\A}=1$ hence $\A=t$, which corresponds to flat (Milne) space, while 
\eqref{A8} then gives the condition on $u$, 
\be                 \label{C0}
0=P_0(u)+u P_1(u). 
\ee
Other possibility to fulfill \eqref{A9} is to set $\A u=F_0=const.$ hence $u=F_0/\A$. 
Equation \eqref{A8} then reduces to 
\be                 \label{deg1}
\frac{3}{\A^2}(\dot{\A}^2-1)=P_0+u P_1=b_0
+\frac{3b_1 F_0}{\A}
+\frac{3b_2 F_0^2}{\A^3}
+\frac{b_3 F_0^3}{\A^3}.
\ee
The four terms on the right here can be viewed as contributions of the graviton interaction terms that mimic a cosmological term,  a gas of domain walls, a gas of cosmic strings, and a dust, respectively. 
This solution is actually known \cite{Chamseddine:2011bu}, \cite{Volkov:2012zb}. 
However,  since $F=F_0$, the reference metric \eqref{BNVf} is degenerate. 

Let us now consider the case where $\cc\neq 1$ and assume first  that $\dot{u}\neq 0$. 
Then \eqref{A7} requires that 
$
2P_1+[\cc(\A u)^{\mbox{.}}-u]dP_1=0.
$
After simple transformations one can show that this condition, together with  \eqref{A6},
are equivalent to the following two conditions:
\bea                      \label{cond}
-\frac{\dot{\A}}{\A}+\frac{1}{\A}+\frac{(c-1) \dot{P}_1}{3P_1}=0, ~~~~~~~~~~~
\frac{\dot{\A}}{\A}+\frac{\dot{u}}{u}+\frac{dP_0}{\cc\A u\, dP_1}=0.
\eea
These conditions can be resolved to algebraically express $\A$ and $\dot{\A}$ in terms of $u$ and $\dot{u}$, 
\be                        \label{AAA}
\A=\A(u,\dot{u}),~~~~~~~\dot{\A}=\dot{\A}(u,\dot{u}).
\ee
Injecting this to \eqref{A8} gives a first order differential equation for $u$,
\be                \label{u2}
\dot{u}^2={\cal Q}(u),
\ee
with a complicated function ${\cal Q}(u)$. In addition, \eqref{AAA} implies that 
\be                   \label{u3}
\frac{\partial \A(u,\dot{u})}{\partial u}\,\dot{u}
+\frac{\partial \A(u,\dot{u})}{\partial \dot{u}}\,\ddot{u}=\dot{\A}(u,\dot{u}),
\ee
which yields a second order differential equation for $u$. Therefore, if $u$ is not constant, it should fulfill 
two differential equations \eqref{u2} and \eqref{u3}. However, \eqref{u2} implies in this case that 
$$
\dot{u}=\pm\sqrt{{\cal Q}(u)},~~~~~\ddot{u}=\frac12\frac{\partial {\cal Q}(u)}{\partial u},
$$
injecting which to \eqref{u3} gives a non-trivial algebraic condition on $u$.
It  follows therefore that $u$ should be constant, hence 
the assumption $\dot{u}\neq 0$ leads to a contradiction. 

Let us therefore return to Eqs.\eqref{cond} and set $\dot{u}=0$. This gives 
\be            \label{C1}
\dot{a}=1,~~~~~~\cc=-\frac{dP_0(u)}{u\,dP_1(u)},
\ee
injecting which to \eqref{A8} leads to 
\be              \label{C2}
P_0(u)\,dP_1(u)-P_1(u)\,dP_0(u)=0.
\ee
These conditions determine values of $u$ and $\cc$, whereas the spacetime metric is again 
flat. 

We note finally that one more possibility to solve Eq.\eqref{A7} is to set $u=0$. Equations 
\eqref{A6}--\eqref{A8} then reduce to 
\be                  \label{deg2}
\frac{3}{\A^2}(\dot{\A}^2-1)=P_0(u),
\ee
however, since $F=\cc\A u=0$, the reference metric is again degenerate. 
Yet one more solution of this type can be obtained 
 by returning to \eqref{A0} and setting there $F=0$.  
It follows then from \eqref{b} that $Y=0$ and $F^2/Y=0$ hence  equations 
\eqref{AA1}--\eqref{AA3} reduce again to \eqref{deg2} with $u$ defined by 
\be              \label{deg3}
u\, dP_0(u)=2u (b_1+b_2\,u)=0. 
\ee
Hence setting $u=-b_1/b_2$ in \eqref{deg2} gives one more solution with $F=0$. The reference metric is again degenerate. 

Summarising, solutions of \eqref{A6}--\eqref{A8} split into two classes. 
First, there are solutions describing a flat Milne spacetime, 
\bea
ds_g^2=-dt^2+t^2\left(dx^2+e^{2x}[dy^2+dz^2]\right),   \nonumber \\
ds_f^2=-(\cc u)^2dt^2+t^2\left((\cc u)^2dx^2+e^{2x}[dy^2+dz^2]\right).
\eea
Here one has either $\cc=1$ while $u$ fulfills  the cubic  equation \eqref{C0} which can have 
up to three real roots,  or $\cc$ is determined by \eqref{C1} while $u$ fulfills the cubic equation \eqref{C2}
which can also have up to three real roots. Therefore, there can be up to six
different values of $\cc u$ and hence six different flat  space solutions.  
These solutions may have different properties.

Other  solutions of \eqref{A6}--\eqref{A8} are of the FLRW type with the scale factor determined 
either by  \eqref{deg1} or by \eqref{deg2}, \eqref{deg3}. However,  one has in this case $F=const.$ hence $dF=0$ 
so that the reference metric is degenerate.

\appendix
\setcounter{section}{1}
\setcounter{equation}{0}
\setcounter{subsection}{0}
\section{Stueckelberg  scalars and new \FFF solutions  \label{AppB}}

\renewcommand{\theequation}{\Alph{section}.\arabic{equation}} 

It turns out that \FFF isotropic solution  \eqref{BNVaa} can be promoted to an 
infinite dimensional family of new solutions. To see this let us first check how this solution
 looks  when expressed in the form similar to \eqref{g2}, \eqref{f2}. 
Making the coordinate shift $x\to x-\chi$ Eq.\eqref{BNVaa} becomes 
\bea                              \label{BNVaab}
ds_g^2&=&-dt^2+\A^2 dx^2+e^{2x }\left[\A^2\left[dy^2+dz^2\right]\right],
 \nonumber \\
ds_f^2&=&-dF^2+F^2\left(dX^2+e^{2 X}[dy^2+dz^2]\right), ~~~~~~~~X=x+f(t)-\chi.
\eea
Combining formulas \eqref{em1}, \eqref{cord1}, \eqref{cord2} one can relate 
the $t,x,y,z$ coordinates 
to coordinates of 5D Minkowski space used in \eqref{g2},
\bea
x^0&=&\frac{\A}{2}\left(e^{-x}+e^x(y^2+z^2+1)\right), ~~~   x^1=\A\, e^x y,~~~~~x^2=\A\, e^x z,~~~\nonumber \\
x^3&=&\frac{\A}{2}\left(e^{-x}+e^x(y^2+z^2-1)\right).
\eea
The inverse transformation is
\bea            \label{BB3}
\A\, e^x=x^0-x^3,~~~~~~y=\frac{x^1}{x^0-x^3},~~~~~~z=\frac{x^2}{x^0-x^3},  \nonumber \\
\A^2=(x^0)^2-(x^1)^2-(x^2)^2-(x^3)^2=(x^4)^2-\frac{1}{H^2}.
\eea
These relations bring  the de Sitter g-metric expressed in the form \eqref{BNVaab} to the 
form \eqref{g2} and back. Similarly, the f-metric in \eqref{BNVaab} 
is transformed to the form \eqref{f2} with 
\bea              \label{BB4}
X^0&=&\frac{F}{2}\left(e^{-X}+e^X(y^2+z^2+1)\right), ~~~   X^1=F\, e^X y,~~~~~X^2=F\, e^X z,~~~\nonumber \\
X^3&=&\frac{F}{2}\left(e^{-X}+e^X(y^2+z^2-1)\right).
\eea
There remains to express these  in terms of $x^0$, \ldots, $x^4$. One has from \eqref{ww2} 
$e^{f-\chi}=u_\ast\,\A/F$ while $F=u_\ast\,\sqrt{w} \A$, hence 
$Fe^X=u_\ast\,\A\, e^x$ and  $Fe^{-X}=u_\ast\,\A\, w\, e^{-x}$. 
Using this and \eqref{BB3} together with \eqref{w2} yields the Stueckelberg fields $X^A$ expressed 
in terms of the 5D Minkowski coordinates, 
\be                                   \label{BB5}
X^0=u_\ast\, \left(x^0+\frac12\,D\right),~~~~X^1=u_\ast\,x^1,~~~~X^2=u_\ast\, x^2,~~~~
X^3=u_\ast\, \left(x^3+\frac12\,D\right),
\ee
where 
\be                         \label{BB6}
D=\frac{(w-1)\,\A^2}{(x^0-x^3)}=\frac{(Hx^4-q^2)^2}{H^2(x^3-x^0)}. 
\ee

Let us 
introduce  lightlike coordinates 
$U=x^3-x^0$ and $V=x^3+x^0$. Then 
the two metrics in \eqref{BNVaab} can be represented as 
\bea
ds_g^2&=&dUdV+(x^1)^2+(x^2)^2+(x^4)^2\,, \nonumber \\
\frac{1}{u_\ast^2}\,ds_f^2&=&dUd(V+D)+(x^1)^2+(x^2)^2+(x^4)^2\,,   \label{BB8}
\eea
where 
\be                     \label{BB9}
UV+(x^1)^2+(x^2)^2+(x^4)^2=\frac{1}{H^2}, 
\ee
and 
\be                         \label{BB10}
D=\frac{(Hx^4-q^2)^2}{H^2 U}. 
\ee
This can be generalised to an infinite dimensional family of new solutions. 
 Specifically, it is known  \cite{Visser} (see also \cite{Mazuet:2015pea})
that if  $P_1=0$ and  $g_{\mu\nu}$ fulfills the Einstein equations with the cosmological 
constant $P_0$ while the two metrics fulfill the Gordon relation,
\be
f_{\mu\nu}=\omega^2\left[g_{\mu\nu}+(1-\zeta^2){\rm V}_\mu{\rm V}_\nu\right],
\ee
where $\omega,\zeta$ are some functions and ${\rm V}_\mu$ is a unit timelike vector,
\be
g^{\mu\nu}{\rm V}_\mu{\rm V}_\nu=-1,
\ee
then the dRGT field equations are satisfied. 
Now, the g-metric in \eqref{BB8} is de Sitter with the Hubble parameter $H^2=P_0(u_\ast)/3$
where $P_1(u_\ast)=0$. Moreover,  
the two metrics in \eqref{BB8} are related to each other via 
\be
ds_g^2=u_\ast^2\left[ds_g^2+dU dD-(dx^4)^2  \right],
\ee
hence the Gordon relation will be fulfilled if
\be                  \label{BB14}
\partial_{(\mu} U\partial_{\nu )}D-\partial_\mu x^4 \partial_\nu x^4=(1-\zeta^2){\rm V}_\mu {\rm V}_\nu\,.
\ee
Let us assume that $D=D(U,x^4)$ and that the vector ${\rm V}_\mu$ has 
non-vanishing components only along the $U$ and $x^4$ directions. 
Then \eqref{BB14} reduce to 
\bea
\partial_U D&=&(1-\zeta^2){\rm V}_U^2 \,, \nonumber \\
\frac12\,\partial_4D&=&(1-\zeta^2){\rm V}_U{\rm V}_4 \,, \nonumber \\
-1&=&(1-\zeta^2){\rm V}_4^2 \,. 
\eea
Taking the square of the second relation and using the two others gives 
\bea
\frac14\,\left(\partial_4D\right)^2=(1-\zeta^2){\rm V}^2_U (1-\zeta^2){\rm V}_4^2=-(1-\zeta^2){\rm V}^2_U
=-\partial_UD,
\eea
hence 
\be            \label{BB17}
\partial_UD+\frac14\,\left(\partial_4D\right)^2=0. 
\ee
Any solution of this PDE provides 
a cosmological solution of the dRGT theory written in the form \eqref{BB8},\eqref{BB9}. 
This gives an infinite dimensional family of new homogeneous and isotropic 
\FFF cosmologies.



\providecommand{\href}[2]{#2}\begingroup\raggedright\endgroup

\end{document}